\documentclass[review,3p,times,11pt]{elsarticle}

\usepackage[version=3]{mhchem}
\usepackage{subfig}
\usepackage{graphicx}
\usepackage[table]{xcolor}
\usepackage{array}
\usepackage{ulem}
\usepackage{float}
\usepackage{amsthm}  
\usepackage{amsmath} 
\usepackage{amssymb} 
\usepackage{mathabx} 
\usepackage{mathtools} 
\usepackage{multirow}
\usepackage{rotating}
\usepackage{textcomp}
\usepackage{float}
\usepackage{url}
\usepackage{color}
\usepackage[colorinlistoftodos]{todonotes}
\usepackage{placeins}
\usepackage{ecrc}
\usepackage{longtable}
\usepackage{booktabs}				
\usepackage{tabularx}
\usepackage{placeins}
\usepackage{lscape}
\usepackage{todonotes} 
\usepackage[separate-uncertainty = true,multi-part-units=single, exponent-product = \cdot, output-product = \cdot]{siunitx}
\usepackage{eurosym}
\usepackage[framemethod=tikz,hidealllines=true,backgroundcolor=blue!20]{mdframed}

\usepackage[final]{changes}
\usepackage{comment}

\usepackage{nomencl}
\makenomenclature

\usepackage{etoolbox}
\renewcommand\nomgroup[1]{%
  \item[\bfseries
  \ifstrequal{#1}{A}{Abbreviations}{%
  \ifstrequal{#1}{G}{Greek symbols}{%
  \ifstrequal{#1}{O}{Other symbols}{}}}%
]}

\volume{00}
\firstpage{1}
\journalname{Journal of Membrane Science}
\runauth{D. Wypysek et al.}
\jid{J. Memb. Sci.}
\jnltitlelogo{J. Memb. Sci.}

\begin{document}
\begin{frontmatter}


\title{How does porosity heterogeneity affect the transport properties of multibore filtration membranes?}

\author[DWI,AVT]{Denis Wypysek}
\author[DWI,AVT]{Deniz Rall}
\author[DWI,AVT,Alberta]{Tobias Neef}
\author[Alberta]{Alex Jarauta} 
\author[Alberta]{Marc Secanell}
\author[DWI,AVT]{Matthias Wessling\corref{mycorrespondingauthor}}

\cortext[mycorrespondingauthor]{Corresponding author: manuscripts.cvt@avt.rwth-aachen.de}

\address[DWI]{DWI - Leibniz Institute for Interactive Materials, Forckenbeckstrasse 50, 52074 Aachen, Germany}
\address[AVT]{RWTH Aachen University, Chemical Process Engineering, Forckenbeckstrasse 51, 52074 Aachen, Germany}
\address[Alberta]{Energy Systems Design Laboratory, Dept. of Mechanical Engineering, University of Alberta, Edmonton, AB, Canada}

\begin{abstract}

The prediction of pressure and flow distributions inside porous membranes is important if the geometry deviates from single-bore tubular geometries. This task remains challenging, especially when considering local porosity variations caused by lumen- and shell-side membrane skins and macro- and micro-void structures, all of them present in multibore membranes. 

This study analyzes pure water forward and reverse permeation and backwashing phenomena for a polymeric multibore membrane with spatially-varying porosity and permeability properties using computational fluid dynamics simulations. The heterogeneity of porosity distribution is experimentally characterized by scanning electron microscopy scans and reconstructed cuboids of X-ray micro-computed tomography scans. The reconstructed cuboids are used to determine porosity, pore size distribution, and intrinsic permeability in the membrane's porous structure in all spatial directions. These position-dependent properties are then applied to porous media flow simulations of the whole membrane domain with different properties for separation layer, support structure, and outside skin layer. Various cases mimicking the pure water permeation, fouling, and backwashing behavior of the membrane are simulated and compared to previously obtained MRI measurements. 

This work reveals (a) anisotropic permeability values and isoporosity in all directions and (b) differing contributions of each lumen channel to the total membrane performance, depending on the membrane-skin's properties. This study encourages to pertain the quest of understanding the interaction of spatially distributed membrane properties and the overall membrane module performance of multibore membranes. 
\end{abstract}
   
\begin{keyword}
Multibore (multichannel) membrane \sep Heterogeneous membrane properties \sep Computational fluid dynamics (CFD) \sep Flow magnetic resonance imaging (MRI)  \sep Backwashing
\end{keyword}


\end{frontmatter}

\sloppy{}

%
%
%
\section{Introduction}

Ultrafiltration is a key technology for the pre-treatment of surface water in water purification plants~\cite{Shannon.2008, Geise.2010, Diop.2002, Singh.2015}. The used ultrafiltration membranes feature a porous structure with spatially varying properties depending on the material and manufacturing method~\cite{Werber.2016}. While pore size distribution and porosity are regarded as homogeneous over the membrane cross-section in ceramic membranes~\cite{Zhu.2015, Zhu.2015b, Zhu.2017, Chi.2017}, polymeric membranes often possess porosity gradients inside the porous structure. Especially in polymeric multibore membranes, these heterogeneities are more predominant than in tubular or fiber membranes because micro- and macro-voids form during fabrication in the precipitation step~\cite{SMOLDERS.1992,Luelf.2018, Spruck.2013, Back.2019}. At the heart of the inhomogeneous porosity is the polymer coagulation process which is highly non-linear in phase separation and solidification and accompanies considerable shrinkage~\cite{Bikel2010}. 

Multibore membranes are increasingly used in industrial processes as they outperform hollow fiber membranes concerning their mechanical sturdiness~\cite{Heijnen.2012, Wang.2014}, which is particularly important during cleaning and backwashing. However, the quantification of their porous structures, their related permeation properties, and their hydrodynamic operation conditions in modules is challenging due to the emerging micro and macro-voids, the presence of porosity gradients in the porous structure, and the incorporation of multiple channels in one porous monolithic structure, each of them influencing each other. Both the formation of the porous structure during the fabrication and the permeation and filtration behavior of multibore membranes~\cite{Wypysek.2021} are subjected to current research.

The performance of multibore membranes is well-studied experimentally~\cite{Gille.2005, Wan.2017, BuRashid.2007, Teoh.2011, SIMKINS.2020}. A majority of these publications focus on the overall performance of the membrane within a membrane module. However, fundamental knowledge of internal flux pathways during permeation, fouling, and backwashing is of great interest. Le Hir et al., for example, used confocal laser scanning microscopy to localize the deposition of fluorescent nanoparticles after the filtration process.~\cite{LeHir.2018} For this technique, however, the authors dried the membranes before the measurement. In contrast, magnetic resonance imaging (MRI) can study internal flux and pathways and fouling phenomena of multibore membranes in-situ and non-invasively~\cite{Schuhmann.2019, Wypysek.2019, Marinkovic2020, Wypysek.2021}. Schuhmann et al. used MRI to investigate the filtration behavior in polymeric multibore membranes. The authors visualized velocity fields in the porous structure of the membrane and gel-layer fouling in the multibore membranes using a sodium alginate solution. They showed uneven flow distribution in the lumen channels.~\cite{Schuhmann.2019} Our recent work~\cite{Wypysek.2019} confirmed and extended these findings by studying pure water dead-end forward filtration and backwashing experiments by combining MRI capabilities with computational fluid dynamics (CFD). Additionally, we performed silica dead-end and cross-flow fouling experiments with subsequent backwashing steps. We found that geometric non-idealities, such as bending of the multibore membrane and the corresponding permeate channel flow patterns, influence hydrodynamic conditions inside the membrane porosity as well as the lumen channels. Furthermore, fouling patterns follow these hydrodynamic conditions.  In our latest work~\cite{Wypysek.2021}, we used MRI to analyze the influence of porosity gradients and several prewetting agents on the initial wetting behavior of multibore membranes. This study showed significant differences in wetting behavior due to the entrapment of the pre-wetting agents in differently sized pores and indicates that microscopic spacial porosity distribution affects macroscopic transport behavior.

Direct measurement of flow distribution inside feed and permeate channels of membrane modules can be investigated via MRI at a macroscopic scale~\cite{Wiese2018}. Yet, quantifying flow through spatially inhomogeneous membranes with porosity gradients at a microscopic scale remains challenging. In our multibore filtration and backwashing study~\cite{Wypysek.2019}, we, therefore, utilized MRI measurements combined with CFD to investigate the non-trivial flow gradients of water permeation and the membrane's behavior during filtration and backwashing. By now, this analysis of flow inside the porous membrane support structure is limited to the rigorous assumption of homogeneous distribution of porosity and permeability. This leads to a bias in the results as the real multibore membrane features porosity gradients at different locations. The research approach presented below addresses a methodology to relieve the previous simplifying assumption of homogeneous porosity and quantify the effect of various porosity distribution features.

Kagramanov et al.~\cite{Kagramanov.2001} analyzed the influence of separation layer and support layer thickness on fluid flow in multibore ceramic membranes with homogeneous pore size distributions in each layer via CFD simulations only. The authors found decreasing filtration efficiency from the center to the periphery depending on the layer thicknesses. Frederic et al.~\cite{Frederic.2018} additionally varied the permeability contrast between the support structure (macroporous region) and the separation layer (microporous region). They found a dependency between the permeability contrast and the channels' contribution to the total permeate flux. However, these distinct regions of separation layer and porous support structure with constant porosities are only valid for ceramic multibore membranes. Due to the manufacturing process of polymeric multibore membranes, these membranes have spatial asymmetries in pore size distribution and porosity. Sorci et al.~\cite{SORCI.2020}, for example, used SEM images to obtain the domain geometry of commercial PES membranes and performed 2D simulations of fluid and particle flow within a small representative microporous structure. In their simulations, they identify large unused regions of the internal pore structure. However, these inhomogeneous support structures with variable spatial properties with gradients are insufficiently studied for the whole membrane cross-section of fibrous polymeric membranes. In general, the design of porosity, surface properties, fluid flow conditions, and their influence on permeation and retention properties remains a challenge. Still, it becomes more and more accessible today through combinations of sophisticated experimental and simulation methodologies~\cite{SORCI.2020, Lohaus2020}. 

\begin{figure}[H]
	\begin{center}
		\includegraphics[width=0.93\textwidth]{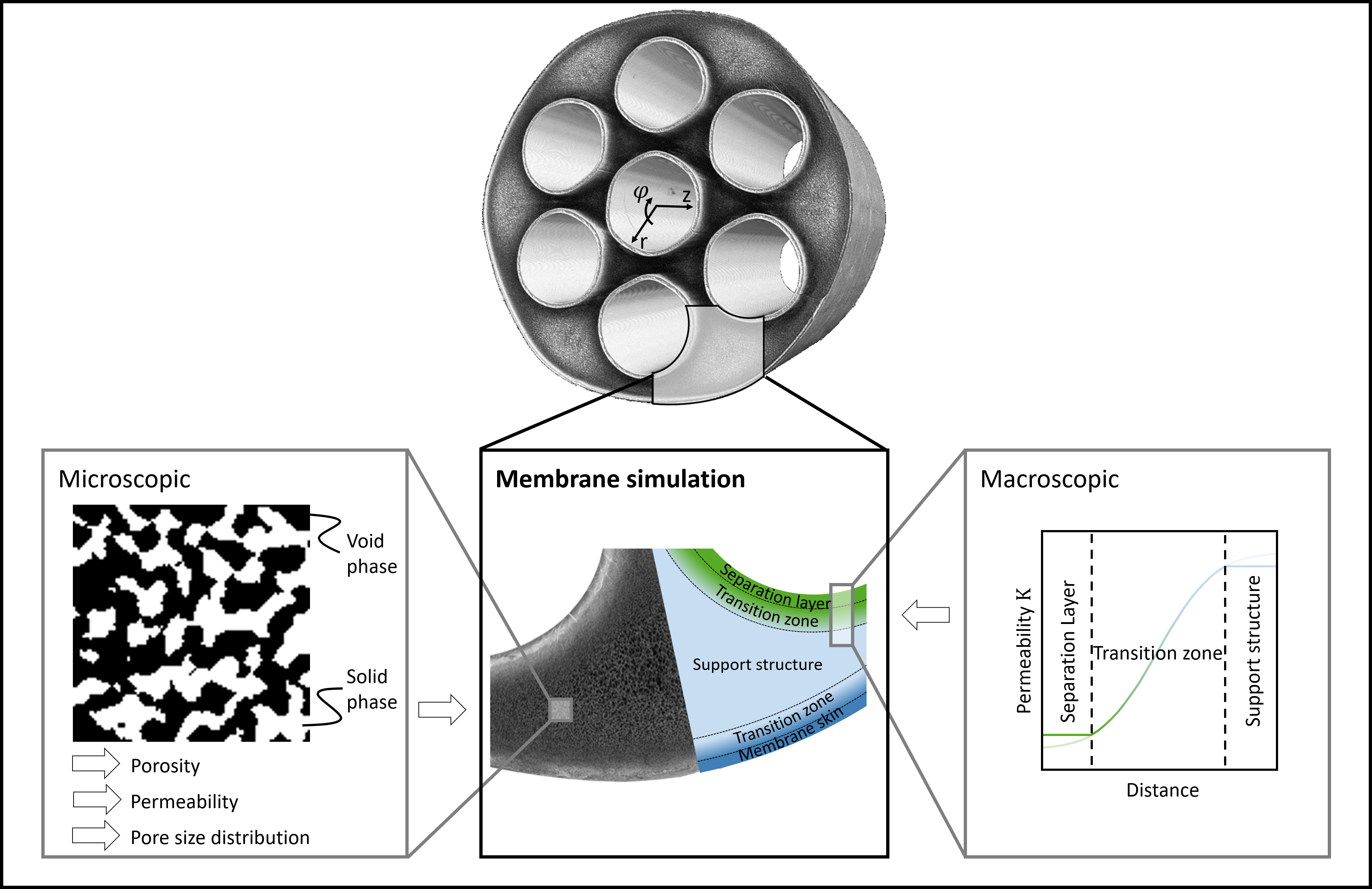}
		\caption[]{Simulation approach - µCT-scans (top) are used to reconstruct position- and direction-dependent samples to obtain microscopic membrane parameters (bottom-left). These membrane specific properties are input parameters for macroscopic, 2D OpenFCST simulations (bottom-middle) with position-dependent properties and gradients (bottom-right).}\label{fig:introduction}
	\end{center}
\end{figure}

This study aims at developing a comprehensive computational fluid dynamics framework to address the missing knowledge of local flow distributions in complex membrane architectures with property gradients. It elucidates hydrodynamic effects like velocity distributions and pressure gradients in a polymeric multibore membrane with heterogeneously distributed material properties during pure water permeation and backwashing mode. We propose the simulation approach illustrated in Figure~\ref{fig:introduction}. 
Using the example of a polymeric multibore membrane, we first analyzed the membrane material and its porosity at different regions and directions (radial, angular, axial) microscopically by scanning electron microscope (SEM) and X-ray micro-computed tomography (µCT) (see Figure~\ref{fig:introduction} top). The obtained images were reconstructed into small cubes for further investigation of their properties and flow pathways. This reconstruction of porous structures based on imaging techniques has recently been established in the fields of geophysics~\cite{Zhao.2018}, fuel cells~\cite{Sabharwal.2018}, and symmetric flat sheet membranes~\cite{Ley.2018}; however, not for polymeric multibore membranes. These reconstructions serve as a volume to apply Stokes flow simulations to obtain values for the intrinsic permeability. 
Second, these reconstructions are utilized to determine membrane-specific, position- and direction-dependent structural properties like porosity and pore size distribution (see Figure~\ref{fig:introduction} bottom-left). 
Third, the obtained structural properties are input parameters for a macroscopic, two-dimensional membrane simulation with the open-source software OpenFCST~\cite{Secanell.2014, Secanell.2017}. 
We developed an OpenFCST code that enables the simulation of gradients and the different zones of the multibore membrane (see Figure~\ref{fig:introduction} bottom-right). Finally, in a case study, the OpenFCST simulation results are compared to the obtained MRI results of our recent study~\cite{Wypysek.2019}. For the latter publication, in fact, the homogeneous porosity assumptions made to simulate the flow distribution inside the multibore failed to predict the measure intra-membrane flow distribution appropriately.

\section{Methodology}
The membrane structure of the multibore membranes (top of Figure~\ref{fig:introduction}) is analyzed and simulated in this work on two different scales: (1) the microscopic scale perspective (cf. Section~\ref{methods_microscopic}) provides insights into the porous structure of the polymeric multibore membrane structure based on µCT-scans and SEM images; (2) the macroscopic scale perspective (cf. Section~\ref{methods_macroscopic}) enables insight into the fluid flow through the membrane as it is installed within a module. The following methodology section presents the analysis and numerical methods used for the two different scale perspectives.

\subsection{Micro-scale property determination}
\label{methods_microscopic}

\begin{figure}[H]
	\begin{center}
		\includegraphics[width=\textwidth]{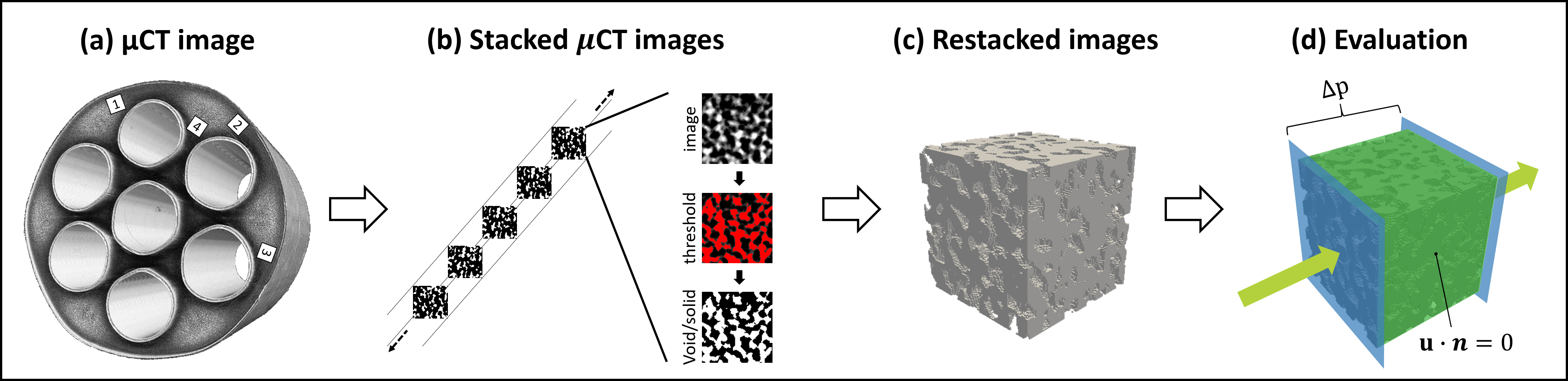}
		\caption[]{Providing insights into the porous membrane material structure through microscopic view: from (a)~a µCT-scan to (b)~a stack of blurred greyscale slices to a stack of black (void phase) and white (solid phase) slices to a reconstructed mesh of the void phase to (c) the restacked cuboid. (d)~Boundary conditions for microscopic Stokes flow simulations for intrinsic permeability calculation.}\label{fig:MethodsMicroscopic}
	\end{center}
\end{figure}

The analysis and reconstruction of the microscopic view of the porous support structure proceed as visualized in Figure~\ref{fig:MethodsMicroscopic}. Reconstructed µCT-scans of a hydrophilyzed PES Sevenbore\texttrademark~fiber from SUEZ SA (Figure~\ref{fig:MethodsMicroscopic}~(a) to (c)) are used to obtain 3D voxel-based meshes of the porous structure for CFD simulations with OpenFCST (Section~\ref{methods_microscopic_reconstruction_CT}). The porosity and pore size distribution of the selected slice and cuboid (Section~\ref{methods_microscopic_porosity_PSD}) are determined. Stokes flow simulations with deal.II~\cite{Bangerth.2016} evaluate the permeability of the support structure based on the µCT-reconstruction of the polymeric matrix (Figure~\ref{fig:MethodsMicroscopic}~(c) and (d)), Section~\ref{methods_microscopic_StokesflowDeal.ii}). As the resolution of the µCT-scans is not sufficient (here \SI{\sim 0.9}{\micro \meter} side length per pixel) to reconstruct the separation layer and membrane-skin layer (pore size of the separation layer according to the manufacturer of \SI{\sim 0.2}{\micro \meter}), we used a resistances-in-series-model (Equation~\ref{eqn:totalkappa}) and experimentally obtained values by Wypysek et al.~\cite{Wypysek.2019} to calculate the intrinsic permeability of the separation layer and the shell-side skin layer (Section~\ref{methods_microscopic_PermeabilityLayer}). SEM images are used to estimate the porosity in these regions.

\subsubsection{Reconstruction of µCT-scans and micro-scale mesh generation}\label{methods_microscopic_reconstruction_CT}
The µCT-image, as shown in Figure~\ref{fig:MethodsMicroscopic}~(a), is obtained by a high-resolution 3D X-ray microscopy device (SkyScan~1272, Bruker). Sample regions 1-4 used for the reconstructions are displayed in Figure~\ref{fig:MethodsMicroscopic}~(a). Samples 1-3 are chosen to evaluate the angular influence of the permeability within the support structure. Sample $4$ is chosen to evaluate the radial dependency of the permeability. Due to the limited space between the single channels, the samples are taken outside of the outer channels. The chosen cuboid samples have a side length of 100 pixels, corresponding to \SI{\sim 90}{\micro\meter}. The reconstructed meshes of the cuboids represent a representative elementary volume (REV) and are sufficient for the sample analysis (according to \cite{GarciaSalaberri.2018, Wargo.2012, Yoon.2013}). Larger sample sizes with homogeneous properties are impracticable due to the small geometry of the multibore membrane. Using the µCT-software \textsc{NRecon Reconstruction}, the REVs are corrected by ring-artifacts reduction, beam-hardening correction, and post-alignment. 

To create the 100$^3$ pixels REV samples, the processed µCT-scan is loaded into the open-source software Fiji~\cite{Schindelin.2012}. The sample regions for samples 1-4 are selected and cropped out of the original stack (100x100 pixels). In the next step, the stack size is set to 100 images (Figure~\ref{fig:MethodsMicroscopic}~(b)). The resulting REVs are then segmented into binary images to distinguish between the void phase and the solid phase of the membrane material using the Otsu method (here, upper slice set to 75).  After applying the threshold, the white color represents the porous media's solid polymer structure, and the black color represents the void phase (Figure~\ref{fig:MethodsMicroscopic}~(b)). 

Subsequently, the binary images are aggregated again to form the reconstructed cuboid. The cuboid is created by transforming the binary images from Fiji to a \textit{vtk}-mesh. The \textit{vtk}-mesh file consists of the spatial location of all nodes, all cells, and their appropriate nodes, as well as all material and boundary identifications (which are explained below). This file is generated through the OpenFCST python environment pyFCST. A script (writeVTK.py) generates a mesh with cells only for the porous structure's void phase. The voxel size in $x$-, $y$-, and $z$-direction is set to the resolution of the µCT-scans (\SI{0.9}{\micro\meter} in each direction). The execution of this script leads to a reconstructed REV, which can be seen in Figure~\ref{fig:MethodsMicroscopic}~(c).

With these REVs, porosities and pore size distributions for the porous support structure can be estimated (Section~\ref{methods_microscopic_porosity_PSD}), and Stokes flow simulations with boundary conditions highlighted in Figure~\ref{fig:MethodsMicroscopic}~(d) can be performed to evaluate intrinsic permeability.

\subsubsection{Analysis of the porosity and the pore size distribution of the support structure}\label{methods_microscopic_porosity_PSD}

The porosity can be assessed directly from the restacked black and white µCT-images. The porosity of a single slice is calculated by dividing the number of black pixels by the total amount of pixels in one slice (\SI{10000}{} in this study). As described by Wargo et al.~\cite{WARGO.2013}, the use of CT imaging is more accurate for determining porosity values compared to SEM imaging.

The pore size distribution (PSD) and the mean pore size diameter are also determined with the restacked cuboids. In this study, the sphere fitting algorithm based on Euclidean distance transform (EDT), described and implemented by Sabharwal et al.~\cite{Sabharwal.2016} (implemented in the open-source writeVTL.py script), was used and is described briefly in Supplement Section~1.

\subsubsection{Governing equations and solution strategy}
\label{ssec:GoverningEquationsMicro}

Fluid flow in the microscale is simulated by solving the steady-state, isothermal, incompressible, single-phase Stokes flow simulations without the influence of gravity written as:
\begin{equation}
    \nabla \cdot \mathbf{u} = 0 \quad \text{in} \quad \Omega
\end{equation}
\begin{equation}
    \nabla \cdot \left( -p \hat{\mathbf{I}} + \hat{\boldsymbol\sigma} \right) = 0 \quad \text{in} \quad \Omega,
\end{equation}
where the shear stress $\hat{\boldsymbol\sigma}$ is given by
\begin{equation}
    \hat{\boldsymbol\sigma} = 2 \mu \nabla_s \mathbf{u},
\end{equation}
and $p$ and $\mathbf{u}$ are the fluid flow pressure and velocity, and $\mu$ the dynamic viscosity in the investigated domain $\Omega$, respectively. The governing Stokes flow equations for mass and momentum balance are explained in more detail in Supplement Section~2. 

The weak form of the equations above is discretized using a Taylor-Hood approximation where linear quadrilateral elements are used to approximate the pressure, and quadratic quadrilateral elements are used for the velocity. The weak form is considered as the Finite Element Method is employed, in which the derivation of the weak form is necessary. This choice stems from the fact that the method is available in OpenFCST, which is intended to solve problems with irregular grids and where several physical phenomena are involved. The resulting system of equations is solved by forming the Schur complement. The entire code for this simulation is a modified version of the open-source step-22 deal.II tutorial~\cite{dealtutorial}. The code is modified to include new boundary conditions, equation parameters, and changing the flow arrangement as explained below. The equations are solved in the computational mesh obtained using pyFCST of the four vtk-meshes of the cuboid reconstructions Figure~\ref{fig:MethodsMicroscopic}~(c)). 

Dirichlet uniform pressure boundary conditions are applied at the surface boundaries in the flow direction (Figure~\ref{fig:MethodsMicroscopic}~(d)). The pressure inlet boundary is set to p~=~\SI{1.0}{\pascal}, while the outlet pressure is set to zero. At all other surface boundaries, a Dirichlet boundary condition with zero normal velocity is imposed. Finally, at internal walls, a no-slip boundary condition is applied.

\subsubsection{Permeability estimation of the support structure} \label{methods_microscopic_StokesflowDeal.ii}
The intrinsic permeability in all spatial directions is computed at post-processing by computing the volumetric flow rate at the inlet surface of the porous media. 
Since we do not expect pore compaction in the applied pressure regimes~\cite{.suez}, Darcy's law (Equation~\ref{eqn:darcypermeability}) is used to calculate the intrinsic permeability of the reconstructed cuboid in flow direction such that, 

\begin{align}
    \kappa=\frac{\dot{V}\mu l}{A \Delta p},
    \label{eqn:darcypermeability}
\end{align}

where $\dot{V}$ is the volume flow that results from the applied pressure difference, $A$ is the cross-sectional area, $l$ is the media's length, $\Delta p$ is the applied pressure difference (set to \SI{1.0}{\pascal} in this study), and $\mu$ is the dynamic viscosity of the fluid (set to \SI{0.001}{\pascal\second} in this study). Length $l$ and cross-sectional area $A$ are based on the cuboid side length. The volume flow is constant in every slice of the mesh along the flow direction. 

To calculate the volume flow (Equation~\ref{eq:VolumeFlow}), the volumetric flux is integrated over the inlet surface, i.e.,
\begin{equation}
   \dot{V} = - \int_A \mathbf{u} \cdot \mathbf{n} \quad  dA 
\label{eq:VolumeFlow}
\end{equation}
where $\mathbf{n}$ is the normal to the inlet boundary, and $A$ is the inlet Area.

\subsubsection{Porosity and permeability estimation for separation layer and membrane-skin interface} \label{methods_microscopic_PermeabilityLayer}

The porosity and intrinsic permeability of the separation layer and the membrane-skin interface cannot be evaluated with the method described above in the case of ultrafiltration membranes. Due to pore sizes below the resolution of the µCT-scans, this imaging method is inappropriate to generate meshes. 

SEM images of the membrane have a higher resolution and can approximate the membrane-skin layer's porous structure and parts of the separation layer~\cite{WARGO.2013} (see SEM images in Supplement Section~13). To obtain a smooth cross-section, the multibore membrane is stored longer than 60 seconds in liquid nitrogen and fractured afterward inside the liquid nitrogen bath. The images are generated by SEM using an acceleration voltage of 5~kV (Hitachi Table Top TM3030 plus). The same cropping and threshold methods are used to obtain the porosity of the separation layer and the membrane-skin layer, as mentioned above. The calculated porosity of the separation layer is overestimated with this method, as SEM images display 3D information in a 2D image. Ten samples are cropped, and the average property values and their standard deviation are computed. FIB-SEM could be used in the future for permeability estimation using the method above~\cite{Sabharwal.2016}.

To estimate the intrinsic permeabilities of the membrane-skin layer and the separation layer, the resistances-in-series model in Equation~\ref{eqn:totalkappa} is used (see~\cite{Singh.2015}). This equation is based on the theory of asymmetric membranes, where the total membrane resistance is the sum of each layer resistance, i.e., 

\begin{equation}\label{eqn:totalkappa}
    \kappa_{\text{total}} = t_{\text{total}} \cdot \left( \frac{t_{\text{support \: structure}}}{\kappa_\text{support \: structure}}+\frac{t_{\text{separation \: layer}}}{\kappa_{\text{separation \: layer}}}+\frac{t_{\text{membrane-skin \: interface}}}{\kappa_{\text{membrane-skin \: interface}}} \right)^{-1},
\end{equation}

where the thickness $t_\text{i}$ of the corresponding domain and the measured total intrinsic permeability $\kappa_{\text{total}}$ is obtained from Wypysek et al.~\cite{Wypysek.2019}. In this study, we assume the membrane-skin layer's intrinsic permeability to be equal to the support structure's intrinsic permeability. This assumption is made because the mean pore radius and the porosity of the membrane-skin layer are closer to the values of the support structure than to the values of the separation layer and is necessary for solving Equation~\ref{eqn:totalkappa} for the unknown separation layer's intrinsic permeability $\kappa_{\text{separation \: layer}}$. As the separation layer is the significant domain for membrane separation, this resistance cannot be neglected.

\subsection{Macro-scale fluid flow simulations}
\label{methods_macroscopic}

The macroscopic simulations provide insights into the fluid flow through the membrane installed within a module. The simulation results on a membrane module scale extend the understanding of experimentally derived polymeric multibore membranes' properties.

\begin{figure}[H]
	\begin{center}
		\includegraphics[width=\textwidth]{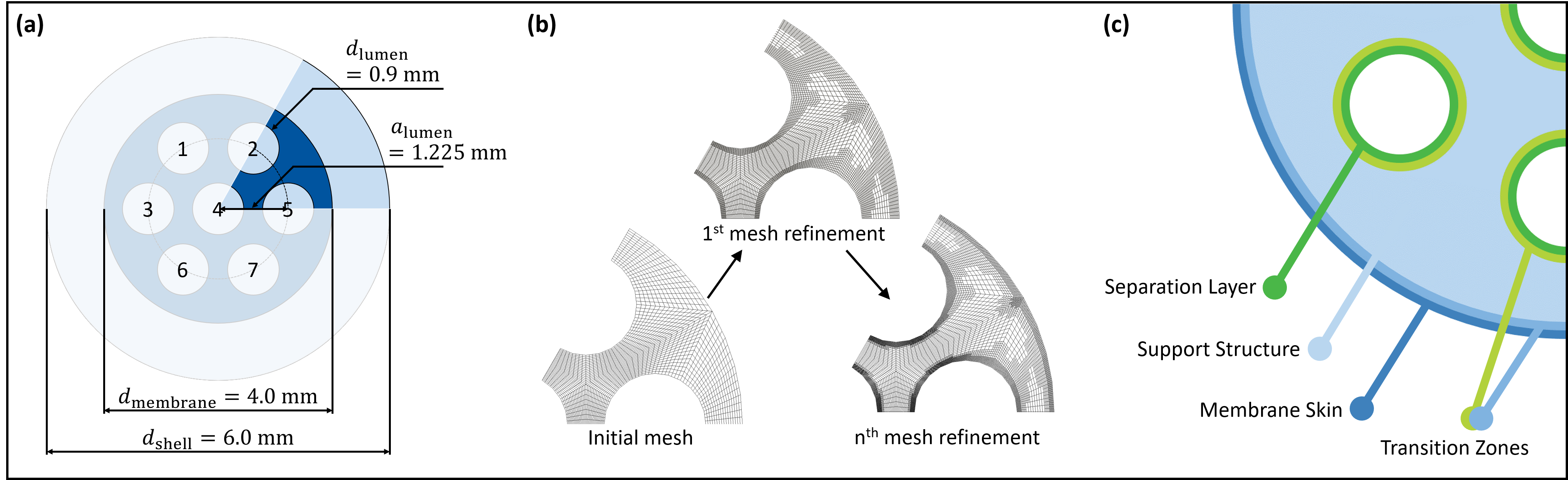}
		\caption[]{(a) Implemented geometry of a multibore membrane with parameters taken from a Sevenbore fiber provided by SUEZ including definition of enumerated lumen channels. (b) Generation of a macroscopic mesh of the membrane geometry, including the subdivision into quadrilateral base elements and adaptive mesh refinements. Elements with the highest result error of the most recent simulation iteration are refined automatically. (c) Abstract layer-based model of the membrane geometry with simplified constant zones for the separation layer, support structure, membrane outer skin, and transition zones guarantees continuity between constant zones and leads to property gradients.}\label{fig:Figure03}
	\end{center}
\end{figure}

The following sections describe methods used to simulate porous multibore membranes. First, the computational domain, generated mesh (see Figure~\ref{fig:Figure03}~(a)) and mesh refinement strategy (see Figure~\ref{fig:Figure03}~(b)) are described in Section~\ref{methods_macroscopic_mesh}. Next, the governing equations of the fluid flow in the porous media and solution strategy are described (Section~\ref{methods_macroscopic_OpenFCST}), followed by the methodology used to implement the position-dependent properties in the porous media (see Figure~\ref{fig:Figure03}~(c), Section~\ref{methods_macroscopic_PositionDependentPropertyMethod}). Finally, the simulation parameters are listed in Section~\ref{methods_macroscopic_setup}. Here, the resulting porous media properties obtained with the methods described in the previous section are assigned as input parameters for the conducted 2D OpenFCST~\cite{Secanell.2014,Secanell.OpenFCSTgithub} simulations. A compressible version of the numerical model was derived and validated by Jarauta et al.\cite{Jarauta.2020} by first solving a lid-driven cavity flow benchmark problem and then simulating a gas permeability experimental setup and comparing experimental and numerical results. In the study at hand, the OpenFCST model is supplemented by a position-dependent property method (see Section \ref{methods_macroscopic_PositionDependentPropertyMethod}), which results are compared to MRI measurements in Section~\ref{sec:comparisonMRI}.

\subsubsection{Mesh generation}
\label{methods_macroscopic_mesh}

A computational mesh with quadrilateral elements is generated to describe the membrane module. For this purpose, the open-source, cross-platform software SALOME 7.3.0~\cite{Ribes.2007} is used. Figure~\ref{fig:Figure03}~(a) depicts the idealized cross-sectional geometry of a membrane module with a multibore membrane in its dry state. The 2D mesh consists only of the membrane domain (dark blue in Figure~\ref{fig:Figure03}~(a)). The geometric values in Figure~\ref{fig:Figure03}~(a) are input parameters for the mesh generation. The mesh consists of five domains: separation layer, support structure, membrane-skin, and two transition zones. Each zone is identified with a different material identifier and will be assigned different properties. Furthermore, the mesh is adaptively refined six times in separation layer and membrane-skin zones to better approximate larger gradients in this region by dividing each quadrilateral element into four equally-sized elements at each refinement level. The used auxiliary geometry with quadrilateral elements as well as the final meshes with magnifications of the boundaries can be seen in Supplement Section~3.

\subsubsection{Governing equations}
\label{methods_macroscopic_OpenFCST}

The steady-state incompressible Navier-Stokes Equations~\ref{eq:NS1} and \ref{eq:NS2} with a friction factor to account for the porous media are solved in the membrane domain, i.e,
\begin{equation}
\nabla \cdot \mathbf{u} = 0 \quad \text{in} \quad \Omega,
\label{eq:NS1}
\end{equation}
\begin{equation}
\rho \nabla \cdot \left( \frac{1}{\epsilon} \mathbf{u} \otimes \mathbf{u} \right) = \nabla \cdot \left( -p \hat{\mathbf{I}} + \hat{\boldsymbol\sigma} \right) +\mathbf{F} \quad \text{in} \quad \Omega,
\label{eq:NS2}
\end{equation}
where
\begin{equation}
\hat{\boldsymbol\sigma} = 2 \mu \nabla_s \mathbf{u},
\end{equation}
\begin{equation}
\mathbf{F} = 
\begin{cases}
\mathbf{0} \quad \text{in} \quad \Omega_c, \\
- \mu \boldsymbol{K}^{-1} \mathbf{u} \quad \text{in} \quad \Omega_p.
\end{cases}
\end{equation}
$\rho$, $p$, $\mathbf{u}$ are the fluid flow density, pressure and velocity, and $\boldsymbol{K}$ is the permeability tensor in the investigated channel domain $\Omega_c$ and porous media domain $\Omega_p$, respectively. A detailed derivation of the volume-averaged form of the Navier-Stokes equations, together with some analysis and validation studies, is given in~\cite{Jarauta.2020} for the compressible form of the equations.

The finite element method is used to solve the system of equations. The non-linear system of equations is solved using a Newton-Raphson algorithm where, at each iteration, the discrete, linearized weak form of the above equations is solved (for an explanation, see Subsection~\ref{ssec:GoverningEquationsMicro}). The solution variables are discretized using a Taylor-Hood approximation, where linear quadrilateral elements are used to approximate the pressure, and quadratic quadrilateral elements are used for the velocity. The resulting system of equations is solved using the direct parallel solver MUMPS. The numerical solver is implemented in OpenFCST~\cite{Secanell.2017}, which uses the parent finite element deal.II libraries~\cite{dealtutorial}. A detailed explanation of the volume-averaging and solution strategy is provided in~\cite{Jarauta.2020} for a compressible version of the same solver, including the solution of some classical benchmark problems such as a lid-driven cavity flow.

\newpage
\subsubsection{Position-dependent property method}
\label{methods_macroscopic_PositionDependentPropertyMethod}

Implementing a continuous position-dependent property method is infeasible with the commercial software COMSOL Multiphysics\textsuperscript{\textregistered} as it can be seen in our previous studies~\cite{Wypysek.2019}. OpenFCST, on the other hand, allowed us to implement a C++ class to impose a position-dependent porosity and intrinsic permeability. In this study, the porosity and intrinsic permeability is a continuous position-dependent function in the whole membrane.

The membrane's geometry is categorized into five zones to account for the different porosity and intrinsic permeability values. Figure~\ref{fig:Figure03}~(c) illustrates the different zones in a sketch of a membrane section. Three zones have constant properties, i.e., the separation layer, the membrane-skin interface, the support structure, and two zones have variable properties allowing for a smooth transition between zones.

While the porosity is a single number, the intrinsic permeability is a rank two tensor. The permeability tensor $\boldsymbol{K}$ contains intrinsic permeabilities in a Cartesian coordinate system. The samples in Section~\ref{methods_microscopic_reconstruction_CT} are cropped and twisted by the angle $\phi$ to account for the cylindrical shape of the membrane fiber. Therefore, the returned $\boldsymbol{K}$-tensor is the product of a base transformation matrix and the evaluated permeabilities from the Stokes flow simulations in section \ref{methods_microscopic_StokesflowDeal.ii}. The procedure of base transformation is explained in Supplement Section~4. 

The values for the constant zones are given by input parameters (summarized in Table~\ref{tbl:openfcstsettingsparam}). The transition zones guarantee a continuous transition of the property values between two neighboring constant property zones. One possibility for such a transition function is the Heaviside function. The original Heaviside function $H$ is a step function with the function values $0$ at $x<0$ and $1$ at $x\geq0$. We used a modified smooth analytical approximation of the Heaviside function (Equation~\ref{eqn:heavysideapprox}), where $K_{max}$ and $K_{min}$ are the property values of the neighboring constant zones, $\Delta x$ is half the transition zone width, $x_0$ is the distance of the middle of the transition zone from the origin, and $q \in [0.5,1]$ is the percentage of the step height $H_{max}-H_{min}$, which is obtained in the distance $\Delta x$ from $x_0$. The distances are obtained by counting pixels in the SEM images of the membrane. Table~\ref{tbl:HeavisideParameters} summarizes input parameters for the used smooth Heaviside function. Figure~\ref{fig:Figure04} illustrates the function and its parameters,

\vspace{5mm}
\begin{minipage}{0.4\textwidth}
\includegraphics[width=1\linewidth,keepaspectratio=true]{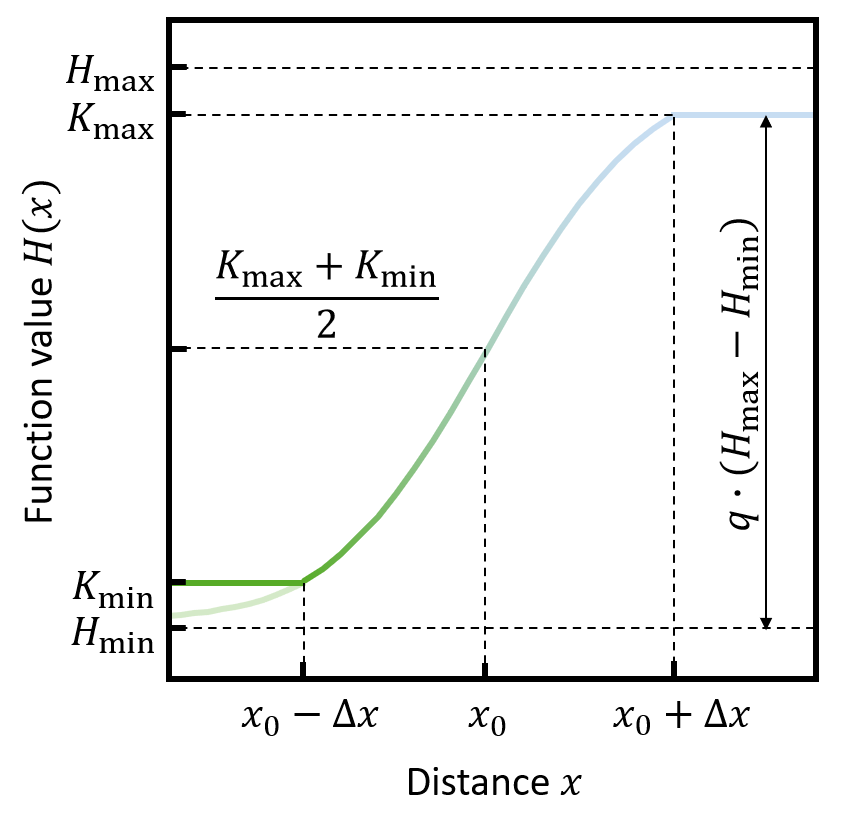}
\end{minipage}
\begin{minipage}{0.55\textwidth}
\vspace*{-10mm}
\begin{equation}\label{eqn:heavysideapprox}
\begin{split}
    H(x) &= H_{min} + \frac{H_{max}-H_{min}}{1+e^{-2k(x-x_{0})}} \\ with \quad k &= \frac{ln(\sqrt{\frac{q}{1-q}})}{\Delta x} \\
    H_{min} &= \frac{K_{max}(1+e^{-2 k \Delta x})-K_{min}(1+e^{2 k \Delta x})}{e^{-2 k \Delta x}-e^{2 k \Delta x}}\\
    H_{max} &= K_{min}(1+e^{2 k \Delta x})-H_{min} e^{2 k \Delta x} 
\end{split}
\end{equation}
\null
\par\xdef\tpd{\the\prevdepth}
\end{minipage}
\captionof{figure}{Smooth Heaviside function used to model the transition zones with a width of $2 \cdot \Delta x$ which guarantees continuous property values between constant zones. The steepness of the transition is regulated with the parameter $q$.} 
    \label{fig:Figure04}
\vspace{5mm}
At each quadrature point in the domain, the Cartesian coordinates of the point are used to calculate its cylindrical coordinates, $r$ and $\phi$, and identify in which zone the point is located. Depending on the assigned zones, the $\boldsymbol{K}$-tensors is calculated. In the three constant zones, the permeability tensor is specified based on the input parameters. In transition zones, the corresponding Heaviside function value determines the $\boldsymbol{K}$-tensor.

\begin{table}[H]
    \centering
    \caption{Input parameters for the smooth Heaviside function obtained from SEM images.}\vspace{1ex}
    \begin{tabular}{ll}\hline
    Parameter&Value\\\hline
    Separation layer thickness  & \SI{55}{\micro\meter}\\
    Skin layer thickness   & \SI{40}{\micro\meter} \vspace{1.5ex}\\
    Separation transition zone width & \SI{55}{\micro\meter}\\
    Skin transition zone width & \SI{40}{\micro\meter} \vspace{1.5ex}\\
    Transition smoothness q [-]&\SI{0.8}{}\\\hline
    \end{tabular}
    \label{tbl:HeavisideParameters}
\end{table}

\subsubsection{Input parameters and boundary conditions}
\label{methods_macroscopic_setup}

Input parameters used for the simulation are summarized in Table~\ref{tab:data.prmParameters}. Six adaptive refinements are used to achieve a grid-independent solution and accurate results where large changes in solution variables exist. The refinement threshold is used to specify the percentage of cells to be refined as only the cells with the highest estimated error are refined. Error estimation is performed using a Kelly error estimator~\cite{Kelly.1983}.

\begin{table}[H]
    \centering
    \caption{Input parameters for OpenFCST simulations.}\vspace{1ex}
    \begin{tabular}{ll}\hline
    Property&Setting\\\hline
    Non-linear solver  & NewtonLineSearch\vspace{1.5ex}\\
    Refinement method & AdaptiveRefinement\\
    Cycles of refinement & 6\\
    Refinement threshold & 0.4 \\
    Coarsening threshold & 0.0\vspace{1.5ex}\\
    Equations & Volume-averaged Navier-Stokes\\
    Simulation conditions & steady-state, incompressible, isothermal, \\
     & single-phase, single-component, no gravity\vspace{1.5ex}\\
    Boundary condition & Outlet: Dirichlet pressure normal stress free at 0 bar\\ 
     & Else: Dirichlet pressure at 0.6 bar\vspace{1.5ex}\\ 
    Fluid & Water at $20^{\circ}C$\\\hline
    \end{tabular}
    \label{tab:data.prmParameters}
\end{table}

Dirichlet pressure boundary conditions are specified in the lumen and shell. For the filtration mode (meaning pure water forward permeation) simulations, high pressure is specified in the lumen and low pressure in the shell. For backwashing mode simulations, the pressure gradient is reversed. A pressure difference of \SI{0.6}{\bar} is specified between lumen and shell based on the pressure results of the simulations from Wypysek et al. \cite{Wypysek.2019}.

To analyze the influence of the membrane-skin interface and to show the possibilities of the developed simulation model, the simulations are conducted under three different conditions:
\begin{enumerate}
    \item The properties of the membrane-skin interface are set equal to the properties of the separation layer (see Section~\ref{sec:OpenFCSTsetting1}. This configuration resembles the multibore membrane's structure best as SEM- and $\mu$CT-scans show lower porosities on the outer skin of the membrane as well as on each lumen channel's surface.) 
    \item The properties of the membrane-skin interface are set equal to the properties of the support structure (see Section~\ref{sec:OpenFCSTsetting2}. With this setting, a membrane without an outer skin is mimicked. Due to the lack of an additional resistance, backwashing efficiency could be improved.)
    \item The properties of the membrane-skin interface are set equal to the properties of the separation layer. An additional damaged zone above channel 1 (top-left) is implemented, which has the same properties as the support structure (see Section~\ref{sec:OpenFCSTsetting3}. During the manufacturing process of multibore membrane modules for MRI analysis, spacers were used to hold the membrane in place, which caused damage to the membrane's outer skin. With this setting, the influence of a damage on the outer membrane skin is analyzed.)
\end{enumerate}

The corresponding permeability and porosity values result from the Stokes flow simulations of the cuboids. These input parameters can be found in Table~\ref{tbl:openfcstsettingsparam}.

\begin{table}[H]
	\begin{center}
		\caption{Parameter configurations for three macroscale simulation settings of a polymeric multibore membrane with OpenFCST. Porosity of separation layer is averaged over 10 SEM images. Permeability of support structure taken from Stokes flow simulations results of reconstructed µCT-images.} \vspace{1ex} 
		\begin{tabular}{llccc}
			\hline
			& & & Settings & \\
						& & 1 & 2 & 3 \\
			&& Membrane-skin =   & Membrane-skin =  & Membrane-skin = \\
			Zone & Variable & Separation layer&Support structure&Separation layer + defect \\
			\hline
			Membrane-skin & $\varepsilon$ [-]            & 0.2 & 0.63 & 0.2 \\
			& $K_{\text{rr}}$ [\SI{}{\square\centi\meter}]       &\SI{1.6e-12}{} & \SI{2.0e-8}{} & \SI{1.6e-12}{} \\
			& $K_{\phi\phi}$ [\SI{}{\square\centi\meter}] & \SI{1.6e-12}{} & \SI{1.3e-8}{} & \SI{1.6e-12}{} \vspace{1.5ex}\\
			
			Separation layer & $\varepsilon$ [-]              & 0.2 & 0.2 & 0.2 \\
			& $K_{\text{rr}}$ [\SI{}{\square\centi\meter}]       & \SI{1.6e-12}{} & \SI{1.6e-12}{} & \SI{1.6e-12}{} \\
			& $K_{\phi\phi}$ [\SI{}{\square\centi\meter}] & \SI{1.6e-12}{} & \SI{1.6e-12}{} & \SI{1.6e-12}{} \vspace{1.5ex}\\
			
			Support structure & $\varepsilon$ [-]         & 0.63 & 0.63 & 0.63 \\
			& $K_{\text{rr}}$ [\SI{}{\square\centi\meter}]       & \SI{2.0e-8}{} & \SI{2.0e-8}{} & \SI{2.0e-8}{} \\
			& $K_{\phi\phi}$ [\SI{}{\square\centi\meter}] & \SI{1.3e-8}{} & \SI{1.3e-8}{} & \SI{1.3e-8}{} \vspace{1.5ex}\\

			Above channel 1  & Defect & false & false & true \\
			\hline
		\end{tabular}
		\label{tbl:openfcstsettingsparam}
	\end{center}
\end{table}

In Section~\ref{sec:comparisonMRI}, OpenFCST simulations are compared to MRI measurement from Wypysek et al.~\cite{Wypysek.2019}. In this simulation, the pressure drop is changed to \SI{50}{\milli\bar} to match the MRI experiments. Setting one is chosen, in which the membrane-skin interface is set equal to the properties of the separation layer because (a) the MRI measurement shows no significant asymmetries in the domain of the membrane, and (b) SEM images (see Figure~\ref{fig:introduction}) reveal the existence of a membrane-skin layer.

In Section~\ref{sec:FoulingComparison}, backwashing experiments after fouling with silica particles are mimicked. To perform this simulation, the blocked domain permeability and porosity are reduced to $\kappa_{\phi\phi}=\kappa_{\text{rr}}=\SI{1.6e-14}{\square\centi\meter}$, and $\epsilon=0.2$. Parameters for start angle of blocking, end angle of blocking, and penetration depth of blocking (which equals a fouling thickness of \SI{10}{\micro\meter}) are also specified. The fouling domain's angles are based on MRI measurements by Wypysek et al.~\cite{Wypysek.2019}. All other parameters are identical to the case of equal membrane-skin and support structure properties. To link all observed phenomena to the influence of the separation layer resistances, a setting without an outer skin layer is chosen. 

For each simulation result, the velocity distribution inside the separation layers of each bore channel was evaluated in a depth of \SI{25}{\micro\meter} from the inner lumen radius and plotted in polar plots. Thus, these plots can be directly linked to each bore channel. A detailed explanation of the method can be found in Supplement Section~5.


\section{Results and discussion}


\subsection{Micro-scale analysis of the reconstructed samples}
\label{sec:redults_microscopic}
The analysis of the four reconstructed µCT-samples provides values for the porosity, the pore size distribution (PSD), and the permeability of the membrane's support structure. First, the porosity is analyzed to ensure homogeneity of the samples in all flow directions to perform post-processing of simulations with Darcy's law. Second, the pore size distribution is analyzed to ensure comparability between the samples and determine the mean pore diameter for the support structure.

\subsubsection{Porosity estimation for the porous support structure using µCT-scans}
Table~\ref{tab:Porosities} lists all layers' average porosities for each evaluated sample and direction with its respective standard deviation. The layer porosity in all three directions fluctuates slightly (standard deviation of maximal 3.3~\%) around the average porosity value. There is also no gradient recognizable from layer one to 100. As an example, the figure of the evaluated porosities of sample 1 plotted against their corresponding image slice can be found in the supplementary material (see Supplement Section~6, Figure~6).

\begin{table}[H]
    \centering
    \caption{Porosity results for each direction for four evaluated µCT-samples, averaged over all layers with corresponding standard deviations.}\vspace{1ex}
    \begin{tabular}{lccc}\hline
    &&Support structure&\\
    &Porosity in $\phi$ [-]&Porosity in $r$ [-]& Porosity in $z$ [-] \\\hline
    Sample 1&0.63 $\pm$ 0.03&0.63 $\pm$ 0.03&0.63 $\pm$ 0.02\\
    Sample 2&0.62 $\pm$ 0.03&0.62 $\pm$ 0.04&0.62 $\pm$ 0.03\\
    Sample 3&0.63 $\pm$ 0.03&0.63 $\pm$ 0.03&0.63 $\pm$ 0.02\\
    Sample 4&0.63 $\pm$ 0.02&0.63 $\pm$ 0.04&0.63 $\pm$ 0.02\\\hline
    Total average&0.63 $\pm$ 0.03&0.63 $\pm$ 0.03&0.63 $\pm$ 0.02\\\hline
    \end{tabular}
    \label{tab:Porosities}
\end{table}

All samples have a similar average porosity from 61.72~\% to 63.39~\% for all directions and a small standard deviation from 1.45~\% to 3.29~\%, which indicates a homogeneous porosity and directional independence for large areas in the membrane's support structure. The smallest average porosity standard deviation is in the extrusion direction of the membrane ($z$-direction), likely due to the manufacturing process, as the membrane is extruded through a nozzle. Given the similar average porosities in all directions, the investigated domain is likely a representative elementary volume. Therefore, it can be used to estimate average transport properties, such as intrinsic permeabilities.

In our previous study~\cite{Wypysek.2019}, we used 2D field emission scanning electron microscope (FeSEM) images to obtain the membrane porosity. One disadvantage of analyzing FeSEM images is the presence of 3D information in a 2D image. That is why FeSEM images do not depict a 2D plane in the membrane correctly. Even if the Sevenbore membranes may vary from batch to batch, the porosity value of 0.82 was overestimated. This limitation is reduced by the µCT-scans in this study. 

\subsubsection{Pore size distribution for the porous support structure using µCT-scans}
The pore size distribution behaves similarly for each sample without significant fluctuations. The mean pore radius is averaged over all mean pore radii and results in \SI{3.7(2)}{\micro\meter}. The support structure's pore size distribution for all four samples is shown in Supplement Section~6, Figure~7).

\subsubsection{Permeability estimation for the porous support structure using Stokes flow simulations}

Intrinsic permeability values for the porous support structure with its respective standard deviations (for all analyzed samples) are obtained from Stokes flow simulations for an applied pressure gradient of \SI{1}{\pascal} and are summarized in Table~\ref{tab:Permeabilities}. 

The results show that the intrinsic permeability values are independent of the angle $\phi$ and radius $r$ of the sample position in the support structure. However, the resulting intrinsic permeability differs depending on the flow direction and suggests an anisotropic support structure. The value in $r$-direction is the highest, and the one in $z$-direction (extrusion direction) the lowest in all samples. This is a new insight from the used multibore membrane. This variation may stem from the precipitation process of the extruded hollow fiber. For subsequent simulation steps, the averages in $r$- and $\phi$-orientation are used as the permeability value. They describe the slice of the membrane simulated by the openFCST framework. 
\begin{table}[H]
    \centering
    \caption{Stokes flow simulation results for intrinsic permeability values in each flow direction for the porous support structure, averaged over four samples with corresponding standard deviations.}\vspace{1ex}
    \begin{tabular}{ll}\hline
    & Support Structure\\
    Direction&Permeability [\SI{}{\square\centi\meter}]\\\hline
    K$_{\text{rr}}$& \SI{2.0 (3)E-8}{}\\
    K$_{\phi\phi}$& \SI{1.3 (2)E-8}{}\\
    K$_{\text{zz}}$& \SI{5.4 (6)E-9}{}\\\hline
    \end{tabular}
    \label{tab:Permeabilities}
\end{table}

The pressure distribution simulation results for the porous support structure for an applied pressure gradient of \SI{1}{\pascal} and the graphs with all resulting intrinsic permeabilities plotted over the four individual samples are shown in Supplement Section~6, Figure~8.

\subsubsection{Porosity and permeability estimations for separation layer and membrane-skin interface}
The estimated averaged mean pore radius, $R_{\text{m}}$, and the averaged porosity, $\epsilon$, of the ten analyzed SEM images of the separation layer, and the membrane-skin interface are listed in Table~\ref{tab:estimation}. The estimated mean pore radius of the separation layer is in good agreement with the manufacturer data of 20~nm~\cite{.suez,suez2}.
\begin{table}[H]
    \centering
    \caption{Results of SEM analysis of mean pore radius and porosity of the separation layer and the membrane-skin interface with corresponding standard deviations. As a comparison corresponding simulated values for the support structure are added.}\vspace{1ex}
    \begin{tabular}{lccc}\hline
    Parameter&Separation layer&Membrane-skin interface&Support structure\\\hline
    $R_{\text{m}}$ [nm]& \SI{18 (2)}{}&\SI{960(40)}{}&\SI{3700 (200)}{}\\
    $\epsilon$ [-]& \SI{0.20 (4)}{}&\SI{0.54 (2)}{}&\SI{0.63 (3)}{}\\\hline
    \end{tabular}
    \label{tab:estimation}
\end{table}
To calculate the separation layer permeability, $K_{\text{ii}}$, the measured total intrinsic permeability of \SI{1.1E-15}{\square\meter} by Wypysek et al.~\cite{Wypysek.2019} is used. We assume the permeability values of the support structure and the membrane-skin interface to be equal as the mean pore radius and the porosity of the membrane-skin interface is closer to the values of the support structure than to the values of the separation layer. This overestimation of the intrinsic permeability of the membrane-skin interface leads to a slightly overestimated intrinsic permeability for the separation layer. Using the resistances-in-series model in Equation~\ref{eqn:totalkappa} and $t_\text{i}$ values obtained from Wypysek et al.~\cite{Wypysek.2019}, the permeability for the separation layer is calculated to: $K_{\text{ii, separation layer}}$~=~\SI{1.6E-12}{\square\centi\meter}. A prediction of the different $K_{\text{ii}}$ values in the separation layer is not possible. Therefore, only one value for all directions is used in the final membrane simulation step.

\subsection{Macro-scale membrane simulation}
\label{sec:redults_macroscopic}
Figure~\ref{fig:Figure05}, Figure~\ref{fig:Figure06}, and Figure~\ref{fig:Figure07} show the predicted velocity and pressure distributions for different membrane-skin interface conditions, illustrated on top of each figure. The top row depicts pure water forward permeation simulation results (from lumen into shell), whereas the bottom row shows backwashing results (from shell into lumen). The left figures provide velocity magnitude data in the corresponding colormap ($\sqrt{v_x^2+v_y^2}$) with normalized arrows depicting the velocity direction. The figures in the middle show the pressure distribution within the membrane domain. The scale ranges from 0.6~bar (set pressure boundary condition) at the inlet boundaries to 0~bar at the outlet boundaries in all Figures. Velocities evaluated in the middle of each separation layer are visualized in polar plots on the right in each figure. 

\subsubsection{Properties of membrane-skin interface set equal to properties of separation layer}
\label{sec:OpenFCSTsetting1}
\begin{figure}[H]
  \centering
  \includegraphics[width=0.99\linewidth]{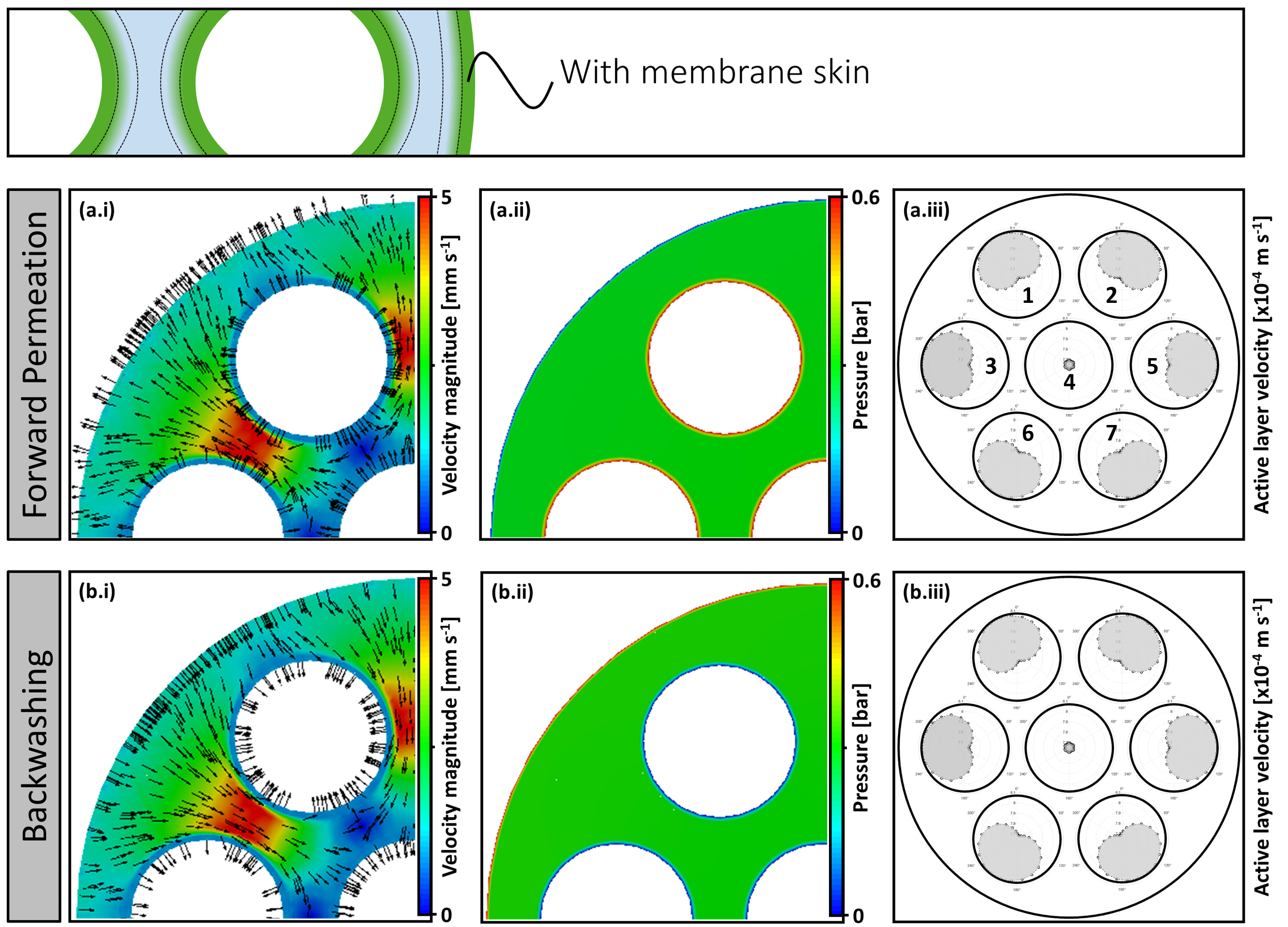}
		\caption{Porous structure velocity ((a.i) and (b.i)), pressure ((a.ii) and (b.ii)) and separation layer velocity ((a.iii) and (b.iii)) distribution inside a polymeric multibore membrane with skin layer properties similar to the separation layer during forward permeation mode (top) and backwashing mode (bottom). Due to rotationally symmetric results, only one-fourth is visualized.}
		\label{fig:Figure05}
\end{figure}

Figure~\ref{fig:Figure05} shows simulation results of a multibore membrane with a membrane-skin interface with the properties of the separation layer. This setting mimics a membrane with a less porous layer on the outer skin and resembles the original membrane best. The scale of the velocity magnitude in this setting ranges from \SI{0}{\milli\meter\per\second} to \SI{5}{\milli\meter\per\second}. It is noticeable that the permeation mode and the backwashing mode are interchangeable in this case. The pressure graphs' colormaps are reversed, the velocity magnitudes are identical, and the arrows for the velocity direction point in the opposite direction. Also, the velocity magnitudes inside the separation layers are equal for each bore, respectively. The behavior can be regarded as ideal and symmetric due to non-existing damage or fouling resistances. Since the results are rotationally symmetric, only one-fourth of the whole membrane cross-section is presented in the flow field and pressure distribution graph. Results for the entire membrane are presented in Supplement Figure~9.

\paragraph{Pressure distribution} The pressure graphs depict a pressure drop in the separation layer (from \SI{0.6}{\bar} to approx. \SI{0.3}{\bar}) as well as in the membrane-skin interface (from approx. \SI{0.3}{\bar} to \SI{0}{\bar}). The pressure gradient in the support structure is negligible compared to the gradients in the separation layer and membrane-skin interface. The total pressure of approx. \SI{0.3}{\bar} in the support structure is the average of the set pressure boundary condition of \SI{0.6}{\bar} at the inlet and \SI{0}{\bar} at the outlet boundary. This value would change with asymmetrical thicknesses of the pressure drop zones. Fluid that enters the support structure will not reenter other bore channels through the high-pressure separation layer because the pressure potential in the support structure is too low. The separation layer and the membrane-skin represent the highest resistances in the membrane. 

\paragraph{Velocity distribution} The velocity graphs also indicate that, for permeation mode, the fluid always follows the path from the lumen into the separation layer into the support structure towards the shell. This flow path is also seen for channel 4 (central). The velocity graphs show a low velocity magnitude in the separation layers and higher magnitudes in the support structure. This velocity distribution is expected based on the higher intrinsic permeability in the support structure. The highest velocity magnitude can be found in the support structure between the outer channels, whereas the lowest magnitude of the velocity is located between channel 4 (central) and outer channels. The high magnitude of the velocity in the narrow region follows the continuity equation for incompressible fluid flow.

\paragraph{Separation layer velocities} It seems that the separation layers have hardly any velocity gradients. However, Figures~\ref{fig:Figure05}~(a.iii) and (b.iii) indicate an uneven velocity distribution around each bore channel's circumference. Here, the velocity plots in the six outer channels are kidney-shaped, with the maximal velocity at the position with the shortest distance to the shell side and the minimal velocity facing towards channel 4 (central). The shorter the distance to the shell side, the lower the resistance for the fluid and thus the higher the velocity in the separation layer (for detailed flow profiles, see Supplement Section~8). The difference between maximal and minimal values in one single bore channel amounts to approx. 4.7~\%. The sum of velocity magnitudes in channel 4 (central) is lower than the magnitudes in the outer lumen channels, with a difference in maximal velocities of approx. 4.6~\%. The lowest velocities in channel 4 (central) are located at the position closest to the outer lumen channels and are identical to the velocity of the outer lumen channel at this position. The highest velocities can be found at the positions between two outer bore channels. However, the difference in maximal and minimal channel 4 (central) velocities amounts to only 0.11~\%.

\paragraph{Mass flux} These unequal velocity magnitudes comparing channel 4 (central) and outer lumen channels lead to different mass flux contributions. The mass flux analysis through the bore channel boundaries shows that the same mass flux exits from every outer lumen channel (14.34~\% of the total mass flux). The mass flux from channel 4 (central) is marginally smaller (13.96~\% of the total mass flux), i.e., less than 3~\% of the mass flux through the outer lumen channel. Consequently, the multibore membrane has the same order of magnitude permeation performance in all its channels. The total mass flux was calculated to
$\dot{m}$~=~\SI{16.89}{\gram\per\second}.

\subsubsection{Properties of membrane-skin interface set equal to properties of support structure}
\label{sec:OpenFCSTsetting2}
\begin{figure}[H]
  \centering
  \includegraphics[width=0.99\linewidth]{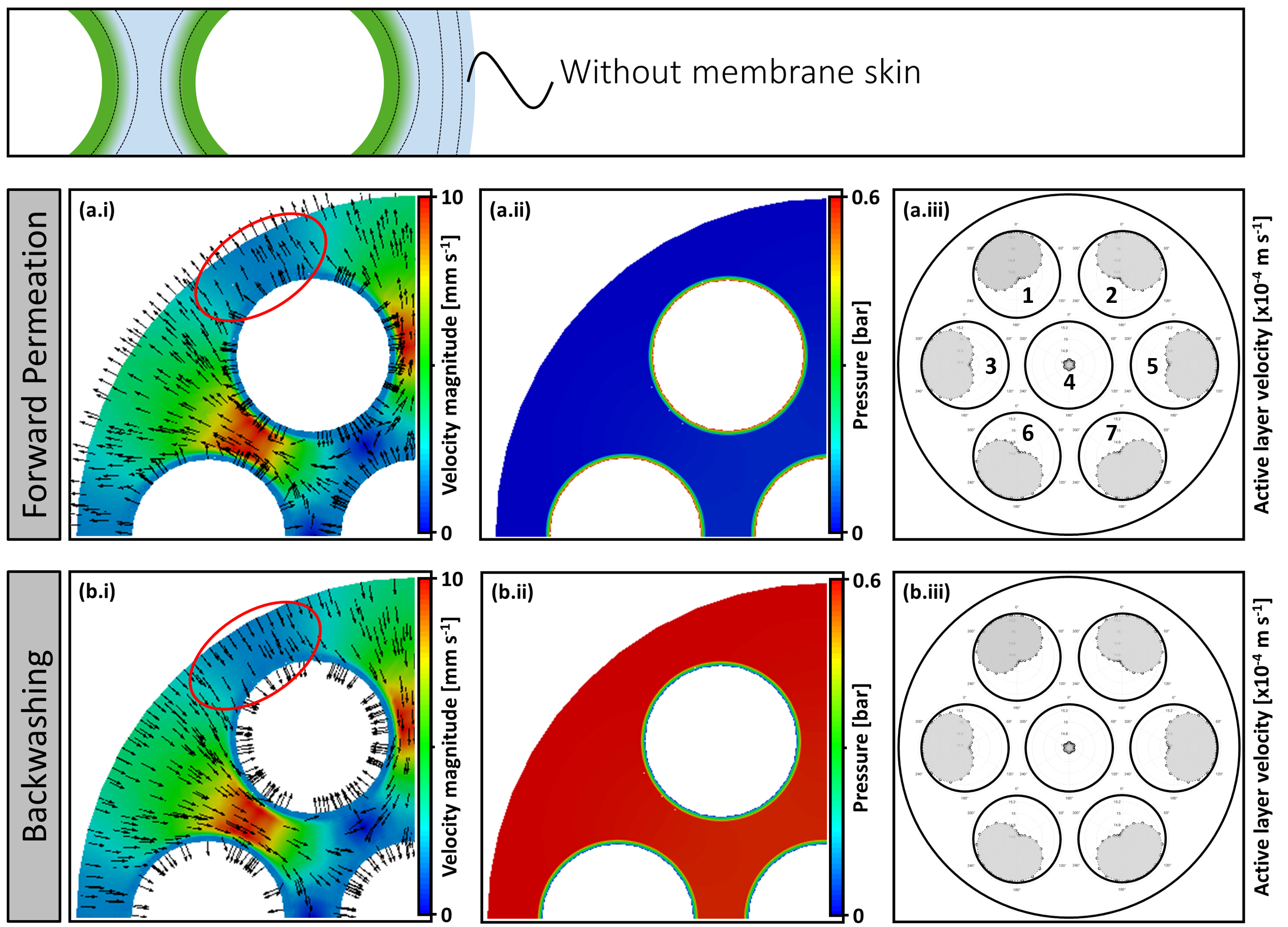}
		\caption{Porous structure velocity ((a.i) and (b.i)), pressure ((a.ii) and (b.ii)) and separation layer velocity ((a.iii) and (b.iii)) distribution inside a polymeric multibore membrane with skin layer properties similar to the support layer during forward permeation mode (top) and backwashing mode (bottom). Red circles mark major differences in fluid distribution to Figure~\ref{fig:Figure05}. Due to rotationally symmetric results, only one-fourth is visualized.}
		\label{fig:Figure06}
\end{figure}

Figure~\ref{fig:Figure06} shows simulation results of a multibore membrane with a membrane-skin interface that has the properties of the support structure. This setting mimics a membrane without a denser layer as the outer skin. In this setting, the maximal velocity magnitude in the scale is approx. twice as high compared to Figure~\ref{fig:Figure05} and ranges from \SI{0}{\milli\meter\per\second} to \SI{10}{\milli\meter\per\second}. The graphs of the permeation and backwashing mode are again interchangeable and rotationally symmetric as well as in Figure~\ref{fig:Figure05}. Results for the entire membrane's cross-section are presented in Supplement Figure~10.
\paragraph{Pressure distribution} Due to the membrane-skin interface setting, this case displays two different zones in the membrane: one with high porosity and intrinsic permeability and one with low porosity and intrinsic permeability. In Figures~\ref{fig:Figure06}~(a.ii) and (b.ii), the main pressure gradient is located in the separation layer around the lumen channels of the membrane. The pressure gradient in the support structure is negligible compared with the pressure gradient in the separation layer. Again, due to the different pressure regimes, the fluid does not reenter other bores.

\paragraph{Velocity distribution} In the velocity plots, similar main conclusions can be drawn as in the results with a membrane-skin in the previous section. One difference is the maximum velocity magnitude that is roughly twice as high. These higher velocities are expected based on the fact that the additional resistance of the outer skin is removed. Hence, the total intrinsic permeability of this setting is higher, leading to higher velocities. Besides the higher velocity magnitudes, the velocity distribution in Figures~\ref{fig:Figure06}~(a.i) and (b.i) is slightly different from those in Figures~\ref{fig:Figure05}~(a.i) and (b.i). Caused by the removed membrane-skin interface, the magnitudes in the area between the outer bore channels and the membrane-skin relative to the maximal magnitude in Figure~\ref{fig:Figure06}~(a.i) and (b.i) are lower (see red circles in Figures~\ref{fig:Figure06}~(a.i) and (b.i)) than the corresponding relation in Figures~\ref{fig:Figure05}~(a.i) and (b.i). The flow distributes more equally if the resistance is higher.

\paragraph{Separation layer velocities} Besides the total velocity magnitudes inside the porous support structure, the separation layer velocities are approx. twice as high compared to the case with a membrane-skin. The major difference is the slightly broader profile close to the shell side. This could be a reason for the lower velocities in the porous support structure in the region between outer bore channels and shell side (red circled area). Otherwise, the velocity profiles in the two cases look similar to each other.

\paragraph{Mass flux} The mass flux analysis through the boundaries shows that the total mass flux is proportional to the total membrane resistance, as expected from Darcy's law. By removing a higher resistance for the membrane-skin interface, the total mass flux increases accordingly. The total mass flux in this setting was calculated to $\dot{m}$~=~\SI{31.57}{\gram\per\second} (Setting~1: $\dot{m}$~=~\SI{16.89}{\gram\per\second}). The difference between mass flux through channel 4 (central) and outer lumen channels remains at 3~\%.

\paragraph{Conclusion} Due to the lack of differences between the two settings in the separation layer, which is responsible for the separation, a membrane with a removed membrane-skin interface performs better regarding the mass flux leaving the membrane. It has a lower total resistance and, therefore, a higher flux for the same applied pressure difference. The results suggest an economically more viable option when considering inside-out filtration mode without an outer membrane-skin. Here additional resistances are mitigated, and backwashing is performed more evenly. 

\subsubsection{Properties of membrane-skin interface set equal to properties of separation layer with damaged zone above channel 1 (top-left) which has properties of support structure}
\label{sec:OpenFCSTsetting3}
\begin{figure}[H]
  \centering
  \includegraphics[width=0.99\linewidth]{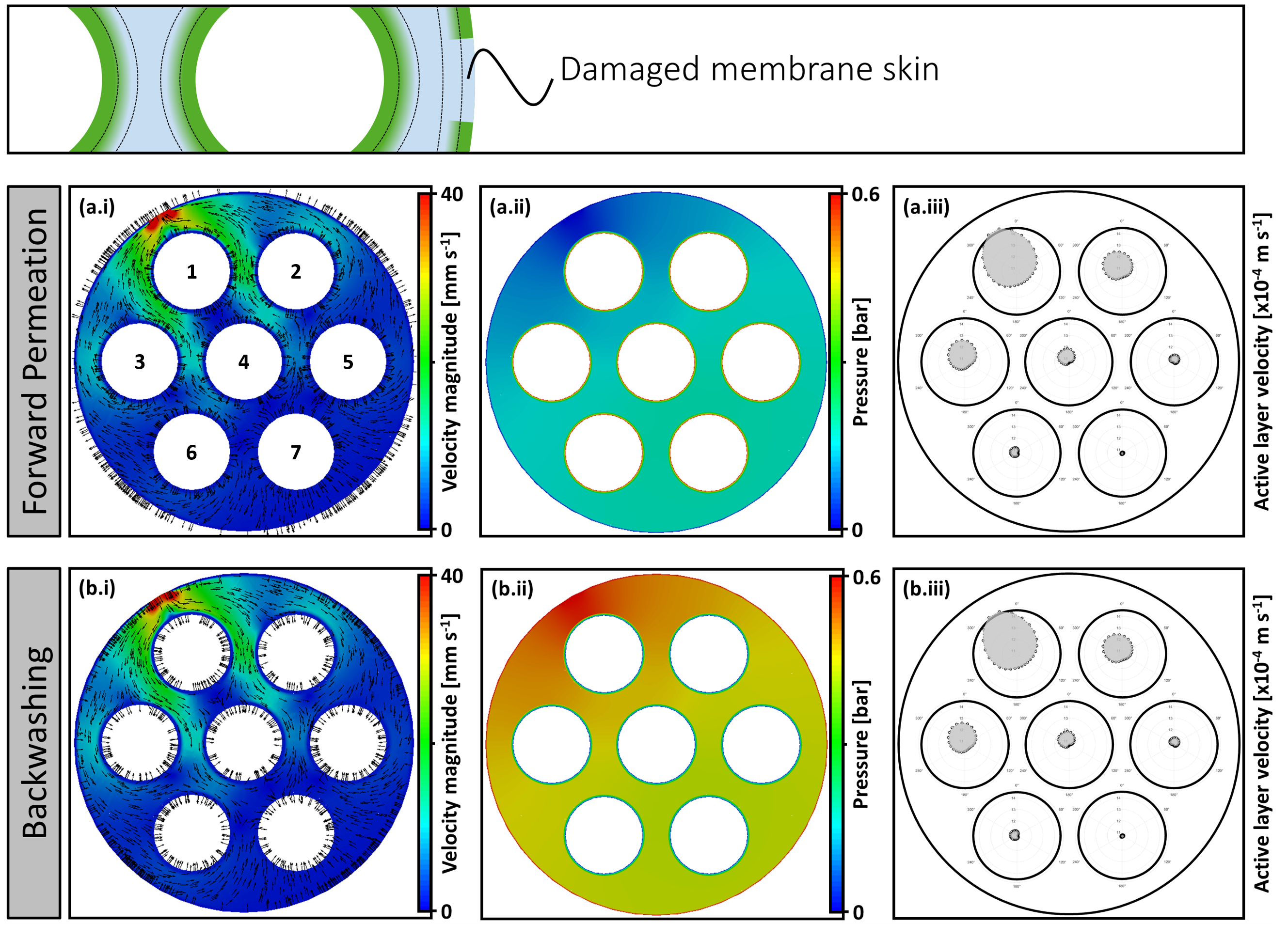}
		\caption{Porous structure velocity ((a.i) and (b.i)), pressure ((a.ii) and (b.ii)) and separation layer velocity ((a.iii) and (b.iii)) distribution inside a polymeric multibore membrane with skin layer properties similar to the separation layer but with a damaged zone above channel 1 (top-left) that has the properties of the support structure during forward permeation mode (top) and backwashing mode (bottom).}
		\label{fig:Figure07}
\end{figure}

Figure~\ref{fig:Figure07} shows simulation results of a multibore membrane with a damaged membrane-skin interface (magnified images can be seen in Supplement Section~9).The membrane-skin interface has the properties of the separation layer, excluding the damaged zone. The damaged zone is located above channel 1 (top-left) of the membrane. It is \SI{0.3}{\milli\meter} wide and has the properties of the support structure. The velocity magnitude ranges from \SI{0}{\milli\meter\per\second} to \SI{40}{\milli\meter\per\second}, which is higher compared to the two simulation results without a damaged zone. The high magnitudes appear in the region around the damaged zone. Hence, the damaged zone significantly changes the flow pattern and pressure distribution in the membrane. Again, the permeation and backwashing modes are the inverse of one another. It has to be mentioned that the flow paths in Figure~\ref{fig:Figure07} result from the position and amount of the observed damages at the outer skin. Changing the position or the number of damaged zones would strongly influence the observed flow field. 

\paragraph{Pressure distribution} The pressure drop in the separation layer and the intact membrane-skin interface is present. However, the pressure drop of the intact membrane-skin interface diminishes as the damaged zone is approached. At this position, the pressure in the damaged zone equals the pressure in the support structure. In this setting, the pressure gradient in the support structure is no longer negligible. The pressure in the support structure rises from the damaged zone towards channel 7 (bottom-right) for permeation mode (and vice versa for backwashing mode). From the bottom-right corner towards the damaged zone, the pressure difference is approx. \SI{0.3}{\bar} for the permeation and backwashing mode. This pressure distribution implies the velocity flow pattern visible in the left column. 

\paragraph{Velocity distribution} The velocity pattern is not rotationally symmetric anymore, as they were in the previous settings. Compared to settings one and two, the damaged zone changes the flow pattern in the whole support structure significantly. The fluid preferably flows through the damaged zone into the shell. Thus, the flow pattern in the porous support structure is directed towards the damaged zone. Nevertheless, the fluid exits (for permeation mode) and enters (for backwashing mode) the membrane's whole circumference. The extreme case of fluid exiting/entering exclusively through the damaged zone is not observed as the pressure gradient boundary condition is valid for the entire circumference. Also, the flow streamlines from the support structure through the intact membrane-skin interface are influenced by the damaged zone. The velocity magnitude is high in the narrow zones between the channel 1 (top-left) and its adjacent outer channels (approx. \SI{20}{\milli\meter\per\second}). Another local velocity hot spot is between the adjacent outer channels and the membrane-skin (\SI{10}{\milli\meter\per\second}) due to the higher radial intrinsic permeability of the support structure. This relation changes for a higher angular intrinsic permeability.

\paragraph{Separation layer velocities} 
The flow profiles inside the separation layers are not symmetrical anymore and are manipulated by the damaged zone. The damaged area has a significant influence on the profiles as they are directed towards it. The velocities on the bore channel's circumference result from an interplay between the distance and position of the damaged zone and the lumen channel's position (for detailed flow profiles, see Supplement Section~10). With increasing distance from the damaged zone, its influence on the flow profile's shape becomes neglectable; however, the velocity magnitudes decrease (max. velocity decrease by 26.7~\% regarding channel 1 (top-left) and channel 7 (bottom-right)). This decreasing behavior can also be seen in the mass flux analysis. 

\paragraph{Mass flux} The mass flux analysis through the separation layers shows that the total mass flux of $\dot{m}$~=~\SI{24.69}{\gram\per\second} is between the two previous settings (Setting~1: $\dot{m}$~=~\SI{16.89}{\gram\per\second}, Setting~2: $\dot{m}$~=~\SI{31.57}{\gram\per\second}). The contribution of the individual channels to the total volume flow through the separation layers in percent can be obtained from Table~\ref{tab:BoreAktivity}. The individual mass fluxes through the channels depend on the distance from the damaged zone. Channel 1 (top-left) has the highest mass flux, followed by channel 2 (top-right) and channel 3 (left), channel 4 (central), channel 5 (right) and channel 6 (bottom-left), and channel 7 (bottom-right). The mass flux of channel 7 (bottom-right) is 16.3~\% lower than the mass flux of channel 1 (top-left). While the damaged zone does not influence the velocity pattern in the separation layer, the mass fluxes are influenced. Ideally, each channel contributes approximately 14~\% to the membrane performance. However, in this setting, channel 7 (bottom-right) contributes only 13.4~\% to the total mass flux.

\begin{table}[H]
    \centering
    \caption{Numerically estimated relative contribution of each individual channel to the total volume flow through the multibore membrane for setting~3 (damage). Numerical values according to simulation results in Figure~\ref{fig:Figure07}.}\vspace{1ex}
    \begin{tabular}{lc}\hline
    Lumen location&Contribution~[\%]\\\hline
    Channel 1 (top-left)&16.0 \\
    Channel 2 (top-right)&14.7 \\
    Channel 3 (left)&14.7 \\
    Channel 4 (central)&13.9 \\
    Channel 5 (right)&13.7 \\
    Channel 6 (bottom-left)&13.7 \\
    Channel 7 (bottom-right)&13.4 \\\hline
    \end{tabular}
    \label{tab:BoreAktivity}
\end{table}

\paragraph{Conclusion} Summarizing, a damage in the membrane-skin interface significantly influences the flow in the whole support structure. This was not visible in our previous study~\cite{Wypysek.2019} as the membrane is modeled with a constant and homogeneous porosity. Thus, the fluid flow inside the porous structure differs from the results in the present study. The results suggest that a membrane with a separation layer on the outside is susceptible to damage, which harms the permeation and filtration result. This clearly illustrates that a membrane without an outer separation layer is also advantageous for inside-out filtration. Damages on the outer surface are buffered by the homogeneous porosity. They are thus mitigating the negative influence on the filtration result. As mentioned above, flow paths and velocities strongly depend on the amount and position of the observed damage. A damaged zone between two outer lumen channels would result in different flow paths compared to Figure~\ref{fig:Figure07} so that flow would be directed towards the damaged zone with higher velocities between those respective lumen channels. How the flow would spread with several positions that mimic damaged zones is outside of the scope of this study and is left as work for future studies.

\subsubsection{Comparison with MRI measurement and old simulation model}
\label{sec:comparisonMRI}
Figure~\ref{fig:Figure08}~(a) depicts the MRI measurement (see magnification in Supplement Section~11), and Figure~\ref{fig:Figure08}~(c) the COMSOL simulation of Wypysek et al.~\cite{Wypysek.2019} for the forward permeation mode. Figure~\ref{fig:Figure08}~(b) shows the resulting graph of a 2D OpenFCST simulation of a multibore membrane with skin layer properties similar to the separation layer properties with a pressure gradient of \SI{50}{\milli\bar} to match the MRI experiment. This pressure difference was chosen to compare flow paths and magnitude. Additionally, in Figures~\ref{fig:Figure08}~(d.1), (d.2), and (d.3), flow profiles taken from MRI measurements and OpenFCST simulations (values taken from between the outer lumen channels, see Figures~\ref{fig:Figure08}~(a) and (b)) are depicted to better compare experiments with simulations. Due to symmetry in the simulation results, respective velocity values between the outer lumen channels are equal to each other.

\begin{figure}[H]
  \centering
  \includegraphics[width=0.99\linewidth]{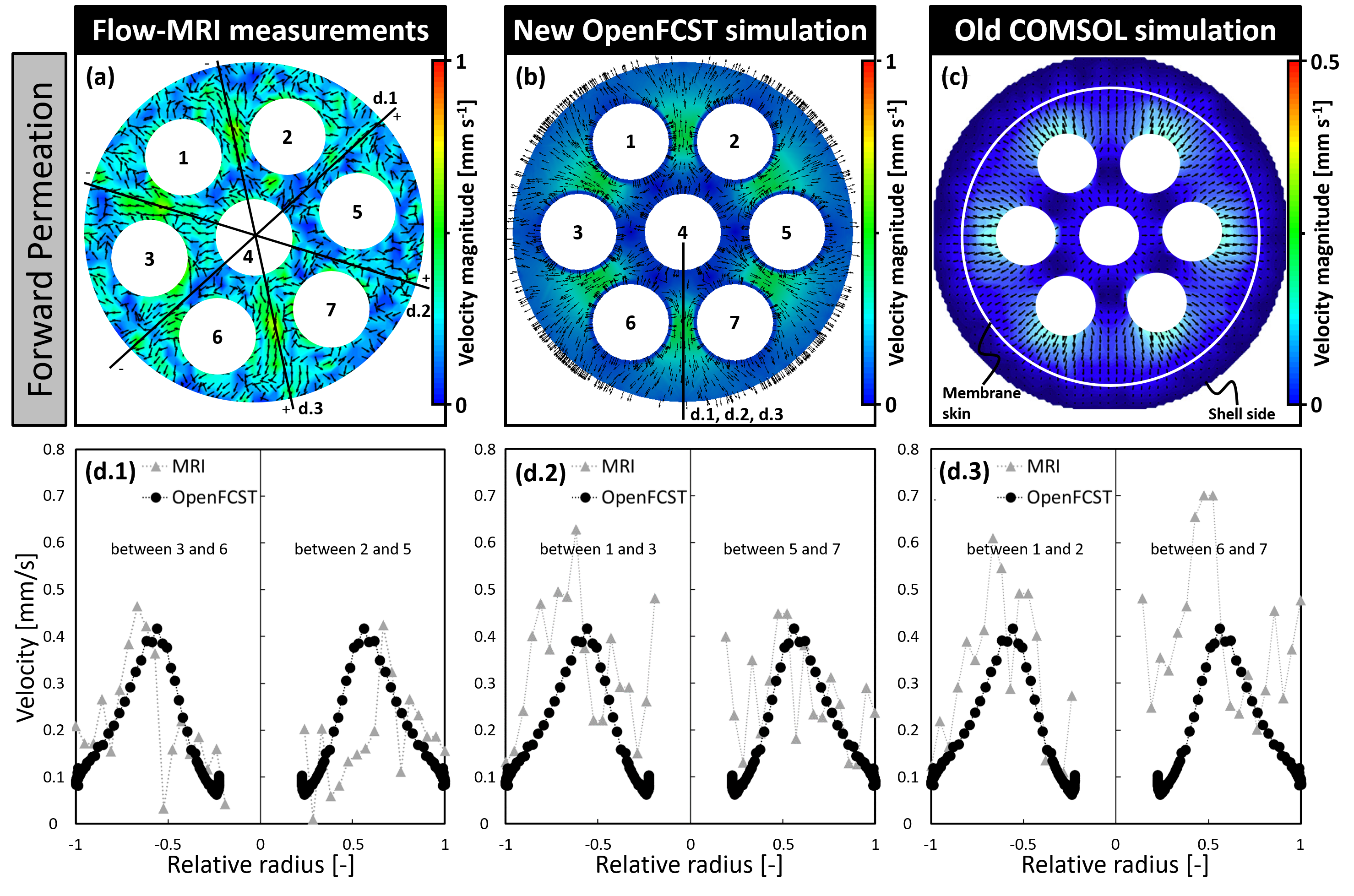}
		\caption{(a) Excerpt of MRI measurement from a whole module's cross-section and (c) COMSOL simulation of radial velocity magnitude in a membrane module with the polymeric multibore membrane in forward permeation mode taken from~\cite{Wypysek.2019}. (b) 2D OpenFCST simulation with radial velocity magnitude of similar order with a total pressure drop of \SI{50}{\milli \bar} of a polymeric multibore membrane with skin layer properties similar to the separation layer in forward permeation mode. Arrows depict the flow direction. (d.1)-(d.3) Velocity profiles from MRI measurements and OpenFSCT simulations taken from between the outer lumen channels for better comparability. MRI and COMSOL simulation image adapted from Wypysek et al. (2019)~\cite{Wypysek.2019} with permission of Elsevier.}
		\label{fig:Figure08}
\end{figure}

Compared to both simulation results (Figure~\ref{fig:Figure08}~(b) and (c)), the MRI measurement (Figure~\ref{fig:Figure08}~(a)) is noisier due to the lower spatial resolution. Nevertheless, regarding the membrane structure and the internal flow pathways, the MRI measurement and the OpenFCST simulation show the highest velocity magnitudes in the narrow zone between the outer channels. In contrast, the COMSOL simulations show the highest velocity magnitudes at the shortest distance between bore channels and membrane skin as the resistance in this direction is the smallest. These phenomena are caused by the application of denser zones in the separation layer and the outer skin. Since in the COMSOL simulation the whole membrane is modeled with equal properties (there is no discrimination between separation layer, porous structure, and outer membrane skin), the velocity pathways differ a lot. Also, channel 4 (central) has hardly any flux in the COMSOL simulation. In contrast, the OpenFCST simulation shows only 3~\% less flow through channel 4 (central) in the ideal case without fouling compared to outer channels. For a better comparison, velocity profiles along the marked lines d.1, d.2, and d.3 (see Figure~\ref{fig:Figure08}~(a) and (b)) can be found in Figures~\ref{fig:Figure08}~(d.1) to (d.3). For all observed positions, the course of MRI measurements resembles OpenFCST simulation results. At the shortest distance between outer lumen channels, the velocity magnitude is highest. The magnitude decreases towards the skin layer and lumen channel 4 (central), at which the velocity increases slightly at the separation layer. 

Depending on the position, velocity magnitude differences can be observed. Low deviations of \SI{1.6}{\percent}, \SI{7.5}{\percent}, and \SI{11,2}{\percent} can be observed for maximum velocity magnitudes between channels 2 and 5, 5 and 7, and 3 and 6, respectively. However, high deviations of \SI{46.1}{\percent}, \SI{50.6}{\percent}, and \SI{67.8}{\percent} can be observed for maximum velocity magnitudes between channels 1 and 2, 1 and 3, and 6 and 7, respectively. These big differences, especially between channels 6 and 7, are caused, on the one hand, by the low spatial and velocity resolution of the MRI measurements~\cite{Wypysek.2019}, and, on the other hand, by MRI measurements of the whole membrane module (including shell side, inlets and outlets). As shown in a previous study~\cite{Wypysek.2019}, secondary flow fields develop on the membrane's shell side, causing drag forces that influence the flow field in the membrane itself. For these reasons, a comparison of velocities between channel 4 (central) and outer channels is challenging, as MRI resolution (especially velocity resolution as velocities are in the range of the noise obtained in the measurements) is too low, and the limit of MRI measurements is reached. However, the course of the velocity profiles of MRI measurements is still comparable to OpenFCST simulation results (see Supplement Section~11).

The same reasons apply to backwashing mode. As secondary flow fields develop on the shell side, backwashing fluid does not enter the membrane equally over the whole circumference~\cite{Wypysek.2019}, making a comparison to our simulation model challenging in backwashing mode. Regarding the simulations with damaged skin (not shown here, see Wypysek et al.~\cite{Wypysek.2019}), the COMSOL simulations show hardly any influence on the flow field inside the membrane structure, while Figure~\ref{fig:Figure07} reveals a huge reorientation of the flow inside the membrane and separation layer towards the damaged zone for the OpenFCST simulation.

In conclusion, the qualitative comparable volume flow rates and, more importantly, similar flow paths through the porous structure in the flow-MRI experiment and the OpenFCST simulation with similar applied pressures show that the developed asymmetric membrane model in this study is more accurate to simulate polymeric membranes than the homogeneous model used in the previous study. COMSOL is only partly suitable for porous structure simulations with gradients in properties. However, the simulation of membranes with spatial equal properties, e.g., ceramic membranes, is possible. Additionally, the flow distribution inside the module (shell side) and the flow inside bore channels are in good agreement with Flow-MRI measurements~\cite{Wypysek.2019}. The previous study's simulation model also allows the simulations of different module configurations (eccentric membrane position, sagging membrane, different outlet positions).

The effect in lumen- and shell-sided velocity and pressure distribution in the MRI measurement and their influence on the flow inside the membrane structure is not assessable with this study and is subject to future studies. The complete membrane module, including gradients in membrane properties, must be considered for the whole picture.

\subsubsection{Example of fouling layer based on previous experiments}
\label{sec:FoulingComparison}
\begin{figure}[H]
  \centering
  \includegraphics[width=\linewidth]{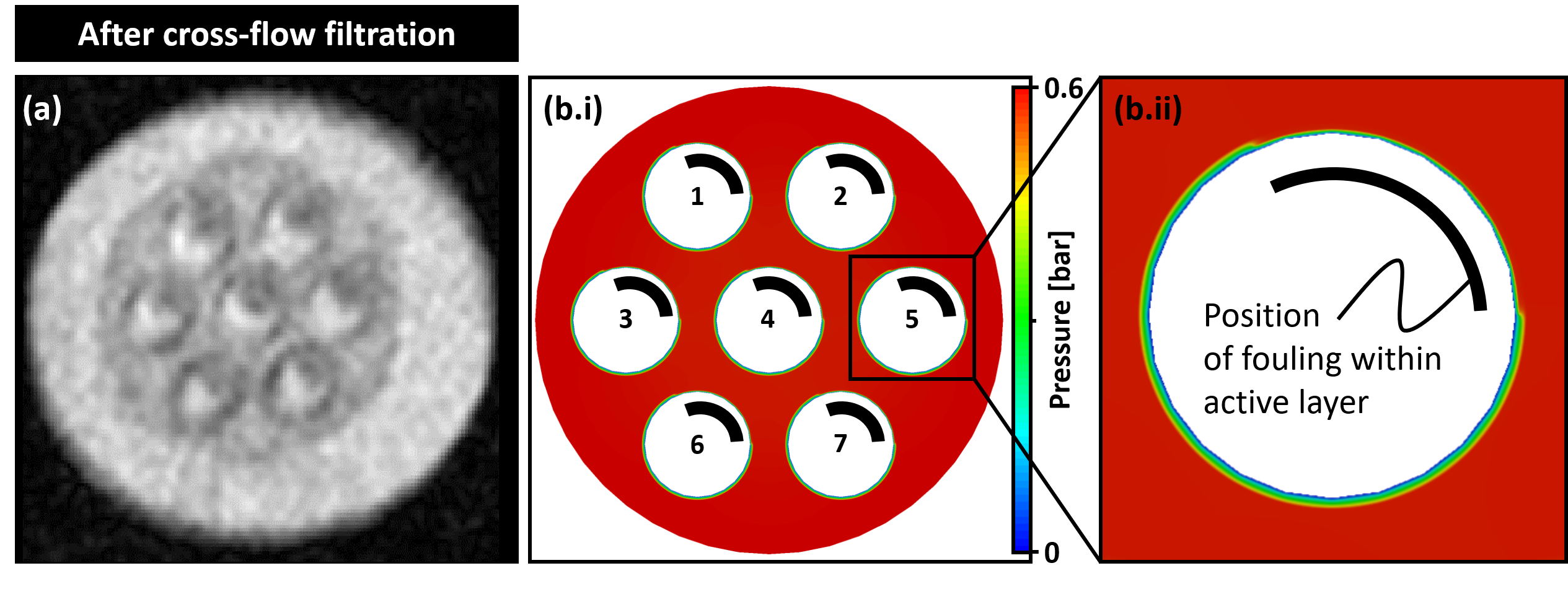}
		\caption{(a) Magnetic resonance image of a whole membrane module after cross-flow fouling and (b.i) simulated pressure distribution with (b.ii) a magnification of a polymeric multibore membrane with skin layer properties similar to the support layer. Black circle sections illustrate the position of fouling in the MRI images. These sections were simulated with a 100 times lower permeability in the separation layer. The pressure difference was set to \SI{0.6}{bar}. Magnetic resonance images taken from Wypysek (2019)~\cite{Wypysek.2019} with permission of Elsevier.}
		\label{fig:Figure09}
\end{figure}

Figure~\ref{fig:Figure09}~(b.i) shows an exemplary pressure simulation result in backwashing mode of a multibore membrane after cross-flow filtration with silica particles (see Figure~\ref{fig:Figure09} (a)). Additional parameters for the blocked domain with less permeability were implemented that mimic the deposition of particles. The position of this domain was taken from MRI measurements (Figure~\ref{fig:Figure09}~(a)), and porosity and permeability values in the separation layer in this region were changed. In this simulation, the membrane-skin interface has the same properties as the porous structure (no membrane-skin layer). Additional pressure distributions results for a multibore membrane before filtration and after dead-end filtration can be obtained from Supplement Section~12. The magnification in Figure~\ref{fig:Figure09}~(b.ii) shows that the pressure drop in the separation layer is not equal over the whole circumference when an additional resistance is present in the respective bore channel. The additional resistance results in a faster pressure drop inside the separation layer. At positions without the fouling resistance, the pressure distribution behaves similarly as in the previous simulations. Neighboring bores do not influence each other. This results in equal hydrodynamic conditions for identical separation layer properties.

\begin{figure}[H]
  \centering
  \includegraphics[width=\linewidth]{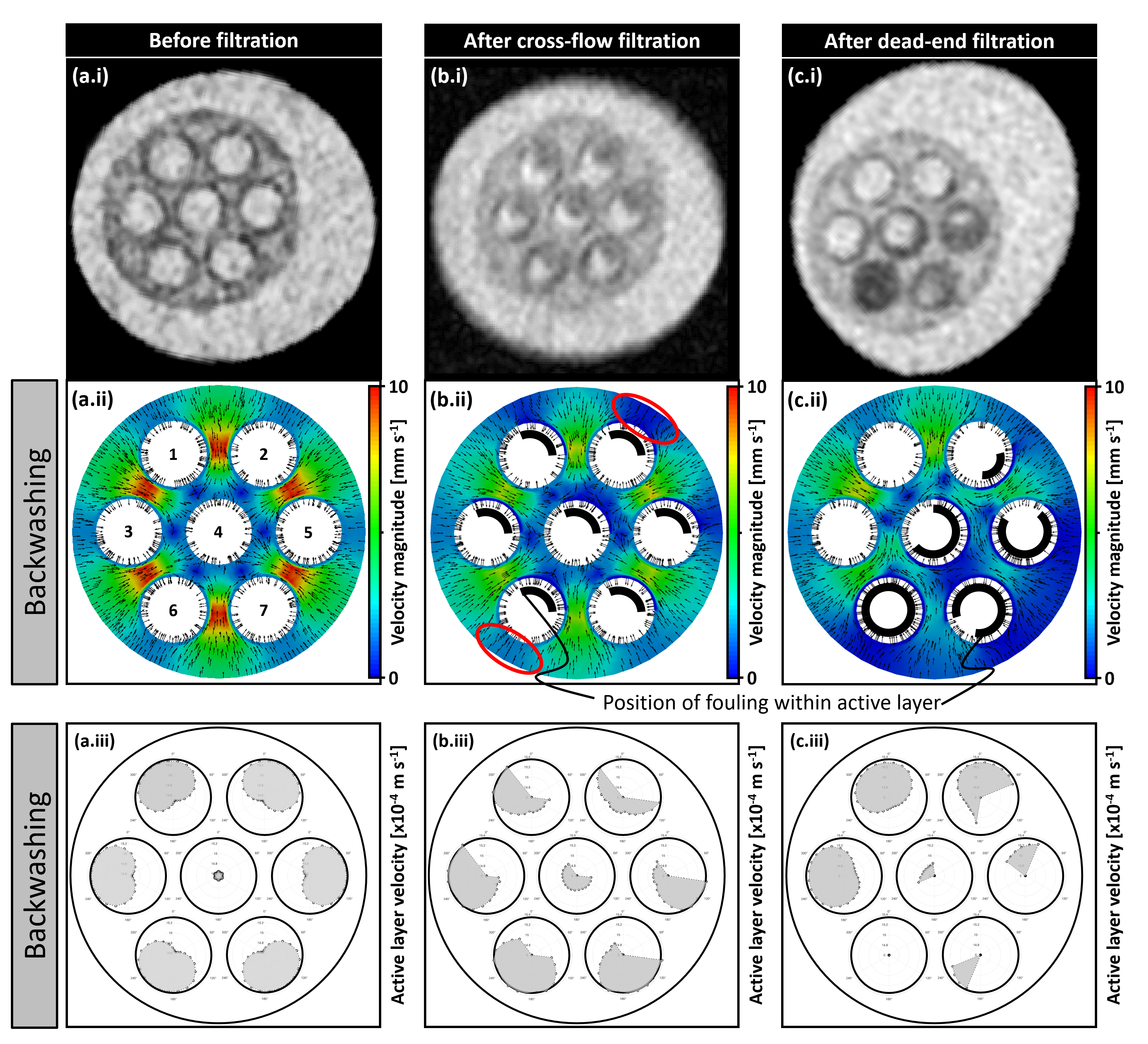}
		\caption{Magnetic resonance images of a whole membrane module (top), simulated porous structure velocity magnitudes with arrows for the flow direction (middle), and separation layer velocities (bottom) in backwashing mode of a polymeric multibore membrane with skin layer properties similar to the support layer in three cases: the membrane is (a) not fouled, (b) fouled during cross-flow filtration and (c) fouled during dead-end filtration. Black circle sections in (b.ii) and (c.ii) illustrate the position of fouling in the MRI images. These sections were simulated with a 100 times lower permeability in the separation layer. The pressure difference was set to \SI{0.6}{bar}. Magnetic resonance images taken from Wypysek (2019)~\cite{Wypysek.2019} with permission of Elsevier.}
		\label{fig:Figure10}
\end{figure}

This behavior can also be seen in the velocity plots in Figure~\ref{fig:Figure10}. Magnetic resonance images are illustrated in the top row and the corresponding simulated velocity maps in the middle row. The bottom row depicts velocity magnitudes in the separation layers. The left column illustrates the reference state, whereas backwashing after cross-flow filtration and after dead-end filtration can be found in the middle and right column, respectively. Velocities in the support structure range from \SI{0}{\milli\meter\per\second} to \SI{10}{\milli\meter\per\second} for all scenarios and up to \SI{1.54}{\milli\meter\per\second} for separation layer velocities. It has to be mentioned that the visualized velocity fields only represent the initial backwashing flow when comparing to our previous study~\cite{Wypysek.2019}. Since fouling layers (depending on the foulant) disappear over time, flow fields would change. This also makes a comparison to flow fields obtained by MRI impossible, as an MRI measurement took approx. 15~min and fouling layers were not visible anymore.

\paragraph{Backwashing reference state} When there is no additional fouling resistance (see Figure~\ref{fig:Figure10}~(a.ii)), the velocity distribution in the membrane is symmetrical. High velocities are located between outer bore channels, low velocities between outer bore channels and channel 4 (central), and outer bore channels and outer shell (see Section~\ref{sec:OpenFCSTsetting2}). Every outer bore channel contributes equally with 14.34~\% to the total mass flow. Channel 4 (central), however, contributes approximately 3~\% less (13.96~\%, see Table~\ref{tab:BoreAktivityFouling} left column). This lower mass flow rate can be disadvantageous for channel's 4 (central) cleaning efficiency as the force to remove particles is lowered. 

\paragraph{Backwashing after cross-flow filtration} If the fouling resistance in all seven bore channels is located at the same position (equal fouling based on fouling volume and location of the foulant after cross-flow filtration, see Figure~\ref{fig:Figure10}~(b.ii)), the velocity distribution is nearly as symmetrical as in the reference state. However, the maximum velocities in the support structure are approximately half for the fouled state. 

Another difference is the uneven distribution in the separation layers (see Figure~\ref{fig:Figure10}~(b.iii)). The velocities at the position of the increased resistance are close to \SI{0}{\milli\meter\per\second}. The fluid needs more force to overcome the fast pressure drop in the high resistance area. It prefers to enter the bore channel through the lower resistance region in the separation layer. The velocity profiles in the separation layers develop through the combination of the bore's location in the membrane and the fouling location. The more the fouling layer is turned away from the shell side (location where liquid enters membrane), the higher the separation layer's overall velocities and vice versa. Also, the profiles are turned towards the shortest distance to the shell side if there is no fouling layer. Channel 4 (central) has the lowest velocities again.

These effects result in a slight velocity maldistribution in the porous structure, which can be seen best when comparing the areas between the outer bore channels and membrane-skin interface in the top-right and bottom-left corner (red circles in Figure~\ref{fig:Figure10}~(b.ii)). In channel 2 (top-right), the area with higher resistance is closer to the membrane-skin interface than to channel 4 (central). Thus, the velocities in the support structure in this region are also close to \SI{0}{\milli\meter\per\second}. The fluid rather flows around the higher resistance area and enters the bore from the other side. In channel 6 (bottom-left), the higher fouling resistance points towards channel 4 (central). Here, the velocity in the area between bore channel and membrane-skin interface is higher (approximately \SI{2}{\milli\meter\per\second}) compared to the velocity at channel 2 (top-right). 

The mass flow contributions range from 14.24~\% to 14.41~\% for the outer channels and 14.09~\% for channel 4 (central), which is only less than 2~\% lower than the outer channels. This deviation is lower as in the reference state, which is advantageous for the cleaning process of the membrane (see Table~\ref{tab:BoreAktivityFouling}. 

\paragraph{Backwashing after dead-end filtration}
In the case of asymmetrical fouling (after dead-end filtration, see Figure~\ref{fig:Figure10}~(c.ii)), the flow distribution in the porous structure is also asymmetric. As the fouling occurs in the bottom-right part of the membrane, the backwashing velocities in this region are smaller compared to the not fouled top-left area. 

The behavior of velocities in the separation layer is similar to the cross-flow case - the profiles depend on the position of the bore and the fouling layer. In the extreme case of complete pore blocking (channel 6 (bottom-left)), the separation layer's whole circumference shows a velocity that is close to zero, which drastically reduces the backwashing efficiency for this bore channel. This can also be seen in the mass flow contributions that are summarized in Table~\ref{tab:BoreAktivityFouling}. 

The completely blocked channel 6 (bottom-left) contributes hardly to the total mass flux in this backwashing mode (0.69~\%). Otherwise, the channels which are not blocked at all (channel 1 (top-left) and channel 3 (left), respectively) have contribution values that are approximately twice as high as in the reference state (28.10~\% and 28.24~\%, respectively). In general, channel 4 (central) contributes less to the total mass flux than the outer channels. Also, the more fouling resistance is added to a bore channel, the smaller the respective mass flux contribution. This uneven flow distribution into each bore channel makes an efficient cleaning challenging. 

In conclusion, dead-end fouling changes the fluid flow distribution between the single bore channels during backwashing. The velocity magnitudes display why the cleaning performance and cleaning times could differ between each channel. Therefore, cross-flow is a measure to prolong the lifetime of the membrane process. 

\begin{table}[H]
    \centering
    \caption{Numerically estimated relative contribution of each individual channel to the total volume flow through the multibore membrane for backwashing in a non-fouled membrane and after fouling in dead-end and cross-flow mode. Numerical values according to simulation results in Figure~\ref{fig:Figure10}.}\vspace{1ex}
    \begin{tabular}{lccc}\hline
    & &Total mass flow contribution&\\
    Lumen location& Reference [\%]& After cross-flow fouling [\%]&After dead-end fouling [\%] \\\hline
    Channel 1 (top-left)        & 14.34 & 14.28 & 28.19 \\
    Channel 2 (top-right)      & 14.34 & 14.24 & 19.48 \\
    Channel 3 (left)                & 14.34 & 14.36 & 28.24 \\
    Channel 4 (central)           & 13.96 & 14.09 & 9.95 \\
    Channel 5 (right)              & 14.34 & 14.27 & 7.94 \\
    Channel 6 (bottom-left)  & 14.34 & 14.41 & 0.69 \\
    Channel 7 (bottom-right)& 14.34 & 14.36 & 5.52 \\\hline
    \end{tabular}
    \label{tab:BoreAktivityFouling}
\end{table}


\section{Conclusion}
This comprehensive computational fluid dynamics study elucidates the hydrodynamic effects in a multibore membrane featuring heterogeneously distributed material properties during permeation and backwashing. It uses spatial-dependent membrane porosities and intrinsic permeabilities in the separation layer, the support structure, and the membrane-skin interface. The transition zones between these layers are modeled by a smooth transition function, representing the property gradients between the zones. Our study closes the gap of measuring macroscopic flow paths inside the membrane and membrane module and modeling flow through the membranes at a microscopic scale with spatial-dependent membrane properties.

First, the spatial-dependent membrane properties (porosity and pore size distribution) are determined using µCT-scans and SEM images of a polymeric multibore membrane. Second, Stokes flow simulations in reconstructed µCT-scan samples are executed and post-processed to obtain permeability estimates. The results of the porous structure simulations with deal.II are required as input to perform OpenFCST simulations of the membrane structure itself. Third, a property function for the multibore membrane geometry is derived and implemented considering the different zones in the membrane. Fourth, the macroscale flow is simulated in the multibore membrane to evaluate the flow pattern within the domain using the Brinkman equations. Thereby, three cases are studied: (1) properties of the membrane-skin interface are set equal to the properties of the separation layer, (2) properties of the membrane-skin interface are set equal to the properties of the support structure, and (3) properties of the membrane-skin interface are set equal to the properties of the separation layer, and an additional damaged zone above channel 1 (top-left) is implemented, which has the same properties as the support structure. Finally, this study's simulation results are compared to the MRI measurements of our previous study~\cite{Wypysek.2019}, and backwashing experiments after fouling with silica particles are mimicked. 

The obtained results provide insights into the membrane's direction-dependent properties and the permeation and backwashing phenomena of polymeric multibore membranes. The porous structure simulations for the micro-scale property determination show that the membrane's support structure in this study is not isotropic. Its radial intrinsic permeability is higher than the angular and axial intrinsic permeability.

The simulations of the entire membrane without additional fouling resistances reveal
\begin{itemize}    
    \item that the main pressure drop is present in the separation layer and/or the membrane-skin interface with similar properties.
    \item that the highest flow velocity magnitudes in the simulations and the MRI measurement are located between the outer channels of the membrane.
    \item that flow patterns and pressure distribution are inversed when comparing forward permeation and backwashing mode without foulant or damaged zone.
    \item that the volume flow through the membrane without a membrane-skin interface is approx. twice as high as compared to with a membrane-skin interface. Hence, membrane permeation and the total flux increase when a skin layer is missing with a missing resistance on the shell side.
    \item that a damaged membrane-skin influences the flow pattern within the support structure. Thus, pressure gradients within the support structure become more pronounced. We hypothesize that flow paths strongly depend on positions and number of damaged zones and the observed flow paths are case-specific.
    \item that each lumen channel contributes nearly equally to the permeation and filtration performance of the membrane. Only marginal differences (3~\%) in the individual contributions are observable. There is no case where a lumen channel contributes less than 13~\% to the total mass flux. However, channel 4 (central) contributes slightly less to the overall mass flux (3~\% less), influencing the cleaning performance. Additionally, the flow distribution within the separation layer depends on the lumen position and is not equal around its circumference.
\end{itemize} 

The simulations of the membrane with additional fouling resistances reveal
\begin{itemize}      
    \item that additional fouling resistances influence pressure and velocity distribution during backwashing mode. These resistances result in a higher pressure drop in the separation layer at positions where fouling occurs. It is assumed that asymmetrical fouling makes recovering each channel equally more difficult.
    \item that the fluid reaches each lumen channel of the membrane, and thus, backwashing is theoretically possible. However, the cleaning performance and cleaning times differ between each channel, as the mass flow contributions differ from channel to channel, and the pathways are different for the outer and the central lumen channels, respectively.
    \item that even a completely blocked bore channel contributes slightly (approx. 0.69~\%) to the total mass flux during backwashing. However, the more a bore channel is blocked, the less it contributes to the total mass flux. Thus, the cleaning performance is different for each channel. For symmetrical fouling, the contribution to the total mass flux is nearly equal in all bore channels.
\end{itemize}

The developed methodology of image-based structure analysis and CFD simulations for asymmetric membranes is essential to understand phenomena within membranes with an asymmetric porosity or permeability distribution. This study can be used as a basis and orientation for future studies with position-dependent properties in asymmetric membranes or membranes with various numbers of lumen channels. The coherence between the measured and simulated velocity magnitudes in the membrane illustrates the applicability of the model. 

Future studies should focus on a whole membrane module (membrane and housing), as our previous study showed an interplay between membrane positioning in the module and the resulting hydrodynamic effects~\cite{Wypysek.2019}. The study at hand emphasizes that the variation of the porosity and the intrinsic permeability is possible. Thereby, other membrane properties can be treated and investigated similarly to the porosity and permeability in the future, for example, the distribution of functional groups of the membrane materials. Furthermore, membranes with several damaged zones at various locations should be studied to understand the interplay between such zones better. Also, a  model for particle transport can gain new insights and lead to a better understanding of fouling physics in hollow fibers and multibore membranes. 
%
%
%
%
%
%
%
%
%
%

\section*{Acknowledgement}
M.W. acknowledges DFG funding through the Gottfried Wilhelm Leibniz Award 2019 (grant ID = WE 4678/12-1) and the support through an Alexander-von-Humboldt Professorship. This project has received funding from the European Research Council (ERC) under the European Union's Horizon 2020 research and innovation program (grant agreement no. 694946). This work was also performed in part at the Center for Chemical Polymer Technology CPT, which is supported by the EU and the federal state of North Rhine-Westphalia (grant no. EFRE 30 00 883 02). This work was enabled by a Bruker SkyScan 1272 funded by the Major Research Instrumentation Program (DFG-Gz: INST 2221157-1 FUGB) as per Art. 91b GG in the Research Building NW1481006 'NGP2 – Center for Next Generation Processes and Products'. T.N. acknowledges financial assistance from the RWTH Aachen - University of Alberta Junior Research Fellowships. M.S. and A.J. acknowledge financial assistance from the Natural Science and Engineering Research Council of Canada (NSERC) grants CRDPJ 501081-16 and RGPIN-2016-04108. The authors thank Karin Faensen for her support in electron microscopy and µCT-measurements.


\bibliographystyle{elsarticle-num}
\biboptions{sort&compress}
\bibliography{Publication1}{}


\setlength{\nomlabelwidth}{2cm}
\setlength{\nomitemsep}{-\parsep}
\nomenclature[A]{MRI}{Magnetic resonance imaging}
\nomenclature[A]{CFD}{Computational fluid dynamics}
\nomenclature[A]{PES}{Polyethersulfone}
\nomenclature[A]{µCT}{X-ray micro-computed tomography}
\nomenclature[A]{OpenFCST}{Open-source Fuel Cell Simulation Toolbox}
\nomenclature[A]{SEM}{Scanning electron microscope}
\nomenclature[A]{FeSEM}{Field emission scanning electron microscope}
\nomenclature[A]{REV}{Representative elementary volume}
\nomenclature[A]{PSD}{Pore size distribution}
\nomenclature[A]{EDT}{Euclidean distance transform}

\nomenclature[O]{$x$}{Axis x-direction}
\nomenclature[O]{$y$}{Axis y-direction}
\nomenclature[O]{$z$}{Axis z-direction}
\nomenclature[O]{$\mathbf{u}$}{Fluid flow velocity}
\nomenclature[O]{$p$}{Fluid flow pressure}
\nomenclature[O]{$\hat{\mathbf{I}}$}{Unit matrix}
\nomenclature[O]{$\dot{V}$}{Volume flow}
\nomenclature[O]{$l$}{Length}
\nomenclature[O]{$q$}{Percentage of step height in Heaviside function, Transition smoothness}
\nomenclature[O]{$H_{max}-H_{min}$}{Step height in Heaviside function}
\nomenclature[O]{$A$}{Cross-sectional area}
\nomenclature[O]{$\Delta p$}{Pressure gradient}
\nomenclature[O]{$\Delta x$}{Half transition zone width in Heaviside function}
\nomenclature[O]{$x_0$}{Distance of the middle of the transition zone from the origin in Heaviside function}
\nomenclature[O]{$\mathbf{n}$}{Normal vector}
\nomenclature[O]{$dA$}{Infinitesimal area}
\nomenclature[O]{$t_\text{i}$}{Thickness of domain i}
\nomenclature[O]{$d_i$}{Diameter of element i}
\nomenclature[O]{$a_\text{lumen}$}{Distance in between two lumen channels}
\nomenclature[O]{$\mathbf{F}$}{Friction factor}
\nomenclature[O]{$H(x)$}{Modified analytical approximation of the Heaviside function}
\nomenclature[O]{$r$}{Cylindrical coordinate in radial direction}
\nomenclature[O]{$R_m$}{Mean pore radius}
\nomenclature[O]{$\dot{m}$}{Total mass flux}

\nomenclature[G]{$\Omega$}{Open connected domain}
\nomenclature[G]{$\Omega_p$}{Porous media domain}
\nomenclature[G]{$\Omega_c$}{Channel domain}
\nomenclature[G]{$\hat{\boldsymbol\sigma}$}{Shear stress tensor}
\nomenclature[G]{$\mu$}{Dynamic viscosity}
\nomenclature[G]{$\kappa$}{Intrinsic permeability}
\nomenclature[G]{$\nabla$}{Nabla operator}
\nomenclature[G]{$\nabla_s$}{Symmetric gradient}
\nomenclature[G]{$\rho$}{Density}
\nomenclature[G]{$\epsilon$}{Porosity}
\nomenclature[G]{$\Phi$}{Cylindrical coordinate in angular direction}

\printnomenclature

\end{document}